\documentstyle[11pt,aaspp4]{article}

\def\beq{\begin{equation}}
\def\eeq{\end{equation}}
\def\beqar{\begin{eqnarray}}
\def\eeqar{\end{eqnarray}}

\def\pref#1{(\ref{#1})}
\def\pcite#1{(\cite{#1})}
\def\etal{{\it et al.~}}

\def\iso#1#2{\hbox{${}^{#2}${\rm #1}}}

\def\li#1{\iso{Li}{#1}}
\def\be#1{\iso{Be}{#1}}
\def\b1#1{\iso{B}{1#1}}
\def\feh{\hbox{$[{\rm Fe/H}]$}}
\def\oh{\hbox{$[{\rm O/H}]$}}
\def\oheq{\hbox{$[{\rm O/H}]_{\rm eq}$}}
\def\dbefe{\hbox{$\omega_{\rm BeFe}$}}

\def\dbfe{\hbox{$\omega_{\rm BFe}$}}

\def\dofe{\hbox{$\omega_{\rm O/Fe}$}}
\def\msol{\hbox{$M_{\odot}$}}
\def\pd{\partial}
\def\nupro{$\nu$-process}
\def\ecr{\varepsilon_{\rm CR}}
\def\decrt{\dot{\cal E}_{\rm CR}}
\def\dnsn{\dot{N}_{\rm SN}}
\def\mejfe{\langle m_{\rm ej,Fe} \rangle}

\def\la{\mathrel{\mathpalette\fun <}}
\def\ga{\mathrel{\mathpalette\fun >}}
\def\fun#1#2{\lower3.6pt\vbox{\baselineskip0pt\lineskip.9pt
  \ialign{$\mathsurround=0pt#1\hfil##\hfil$\crcr#2\crcr\sim\crcr}}}

\def\bline{\rule[1.2mm]{3em}{0.1mm}}

\begin{document}
\rightline{UMN-TH-1826/99}
\rightline{TPI-MINN-99-51}
\rightline{astro-ph/9911320}
\rightline{November 1999}

\title{TESTING SPALLATION PROCESSES WITH BERYLLIUM AND BORON}

\author {Brian D. Fields}
\affil {Department of Astronomy\\ University of Illinois\\
Urbana, IL 61801, USA}

\author{Keith A. Olive}
\affil{Theoretical Physics Institute\\ School of Physics and Astronomy\\
University of Minnesota\\ Minneapolis, MN 55455, USA}

\author{Elisabeth Vangioni-Flam}
\affil{Institut d'Astrophysique, 98 bis Boulevard Arago \\
Paris 75014, France}

\author{Michel Cass\'{e}}
\affil{Service d'Astrophysique, CEA, Orme des Merisiers\\
  91191 Gif sur Yvette, France\\
 and\\
 Institut d'Astrophysique, 98 bis Boulevard Arago \\
Paris 75014, France \\
 }
            
\begin{abstract} 
The nucleosynthesis of Be and B by spallation processes
provides unique insight into the origin of cosmic rays.  
Namely, different spallation schemes predict sharply different trends
for the growth of LiBeB abundances with respect to oxygen.
``Primary'' mechanisms predict BeB $\propto$ O, and are well motivated by
the data if O/Fe is constant at low metallicity. In contrast,
``secondary''  mechanisms predict BeB
$\propto$ O$^2$ and are consistent with the data if O/Fe increases
towards
low metallicity as some recent data suggest. Clearly, any primary
mechanism, if operative, will dominate early in the history of the Galaxy.
In this paper, we fit the BeB data to a two-component scheme which includes
both primary and secondary trends.
In this way, the data can be used to probe the period in which primary
mechanisms are effective. We analyze the data using consistent stellar
atmospheric parameters based on Balmer line data and 
the continuum infrared flux.   
Results
depend sensitively on Pop II O abundances and, unfortunately, on the choice
of stellar parameters.  When using recent results which show O/Fe
increasing toward lower metallicity,  a two-component Be-O fits indicates
that primary and secondary components contribute equally at 
$\oh_{\rm eq} = -1.8$ for Balmer line data; and $\oh_{\rm eq} = -1.4$ to
$-1.8$ for IRFM. We apply these constraints to 
recent models for LiBeB origin.
The Balmer line data does not show any evidence for primary production.
On the other hand, the IRFM data does indicate a preference
for a two-component
model, such as a combination of 
standard GCR and metal-enriched particles accelerated in superbubbles.
These conclusions rely on a detailed understanding of the 
abundance data including systematic effects which may alter the derived
O-Fe and BeB-Fe relations.

\end{abstract}

\newpage

\section{Introduction}

The light elements beryllium and boron
(BeB)
provide unique insight into the nature of
non-thermal nucleosynthesis
in our Galaxy.
Due to their low binding energy, these nuclei
are not produced significantly in the big bang
(Thomas, Schramm, Olive, \& Fields \cite{tsof}, Delbourgo-Salvador \&
Vangioni-Flam \cite{dsvf}) or in stellar nuclear burning.
Instead, BeB are made by spallation processes
due to energetic nuclei and neutrinos. 
The nucleosynthesis origin of BeB is in principle 
encoded in their
abundances in the most primitive, metal poor (Population II)
stars.  Indeed, the mere presence of Be in 
halo stars is perhaps the strongest evidence that accelerated
particles
were present in the early Galaxy.\footnote{Pop II lithium abundances are
dominated by the primordial \li7 component, which must be subtracted
to provide information about 
Galactic Li sources.  
Elemental Li is discussed in
Ryan, Beers, Olive, Fields, \& Norris \pcite{rbofn};
the Li isotopes are analyzed in
Vangioni-Flam \etal \pcite{evf3} and
Fields \& Olive \pcite{fo99b}.}
Thus, in this paper we will focus on Pop II BeB trends,
via a detailed analysis of the data and 
a comparison with recent models.

The production of BeB by Galactic cosmic ray (GCR) spallation
of interstellar CNO nuclei was the standard model for
BeB nucleosynthesis for almost two decades after
first being proposed (Reeves, Fowler, \& Hoyle \cite{rfh};
Meneguzzi, Audouze, \& Reeves \cite{mar}).
However, this simple model was challenged by the
observations of BeB abundances in Pop II stars,
and particularly the BeB trends versus metallicity.
Measurements showed that both
Be and B vary roughly {\em linearly} with Fe,
a so-called ``primary'' scaling.
In contrast,
standard GCR nucleosynthesis predicts that BeB should be 
``secondary'' versus spallation targets CNO,
giving $\be{} \propto {\rm O}^2$ 
(Vangioni-Flam, Cass\'e, Audouze, \& Oberto \cite{evf1}).
If O and Fe are co-produced (i.e., if O/Fe is constant)
then the data clearly contradicts the canonical theory, i.e. BeB production via standard GCR's.

BeB abundances in Pop II stars have thus provided the motivation
for primary BeB nucleosynthesis, in which the
production rate is independent of metallicity.
These primary models invoke the extraction and acceleration of fresh products
of nucleosynthesis ($\alpha$'s, CO) originating from Type II supernovae 
and Wolf-Rayet stars and 
fragmenting on interstellar H and He. They should be 
considered in addition to the standard GCR (secondary) process; any
primary process would be expected to play a major role in the early Galaxy
(halo phase), whereas the secondary one takes over in the galactic disk
(Cass\'e \etal   1995, Vangioni-Flam \etal 1996) or earlier as we shall
see.
 The transition between the two modes may vary depending on 
the adopted BeB data, as well as the [O/Fe] vs [Fe/H] correlation.

One such primary model has focused on
superbubbles--regions of hot, rarefied, metal-rich gas 
swept out by the collective effects
of massive star winds and supernova explosions.
These regions have been proposed 
(Vangioni-Flam et al.\ \cite{vroc};
Higdon \etal \cite{hlr};
Ramaty \& Lingenfelter \cite{rl};
Parizot \cite{par}; 
Vangioni-Flam, Cass\'e
\& Audouze 1999) 
to accelerate freshly synthesized, metal-rich particles.
The energy spectrum of these nuclei
has been extensively
studied by Bykov (1999,\cite{bykov})
and
it is remarkably similar to the GCR injection spectrum.
Finally, core collapse supernovae themselves have been proposed
as a site of \b11 production (and some \li7) via
spallation reactions between
supernova neutrinos passing through the \iso{C}{12} layer,
the ``$\nu$-process'' (Woosley \etal \cite{woo1}).

Recently, another solution has been proposed
to resolve discrepancy between the observed BeB
abundances as a function of metallicity and the predicted secondary trend
of GCR spallation. As noted above, standard GCR nucleosynthesis
predicts $\be{}\propto {\rm O}^2$, while
observations show $\be{} \sim {\rm Fe}$, roughly; 
these two trends are inconsistent if O/Fe is constant
in Pop II.  However, recent
observations find O/Fe increasing at low metallicities
(Israelian, Garc\'{\i}a-L\'{o}pez, \& Rebolo \cite{igr}; 
Boesgaard \etal \cite{boes}). In fact Boesgaard \etal
\pcite{boes} argue for a single slope of $\simeq -0.35$ for O/Fe
vs. Fe {\em at all} metallicities.  As shown by Fields \& Olive
\pcite{fo99a}, a combination of standard GCR nucleosynthesis, and
$\nu$-process production of \b11 is consistent with current data. 
One should note that while the trend in O/Fe is seen in the OH data
and agrees with the O I triplet (Boesgaard \etal \cite{boes}), they
continue to disagree with the observations of O/H using the forbidden
[O I] line (Fulbright \& Kraft 1999).

It is clear that spallation processes, BeB abundances, and O-Fe
evolution are closely linked.  
In this paper, we explore these connections via
careful analysis of both data and theory.
We use the most current BeB data,
with abundances derived from consistent stellar atmosphere
models.  We consider stellar parameters based the Balmer-line data of
Axer, Fuhrmann, \& Gehren \pcite{afg} and the IRFM (Alonso \etal
\cite{aam2}). We first fit the data to both iron and oxygen to
determine the overall primary vs. secondary nature of the BeB
abundances.  
These data are fit to a
two-component, primary plus secondary, metallicity dependence. This fit
quantifies both the strength of each component as well as the metallicity
and hence epoch as which  the secondary component becomes dominant. 

We compare these phenomenological results to the predictions of current
models for  BeB nucleosynthesis and chemical evolution. We find that for
the Balmer line data, any primary component could only have been dominant
at oxygen abundances [O/H] $< -1.8$, which is at metallicities below the
existing data.  That is, for this data set, we find no evidence for
primary BeB production.  In contrast, for the IRFM data, we find that 
primary production dominates for [O/H] $< -1.4$ to $-1.8$ (depending
on the exact data-see below).  The higher value indicates
indicating the need for a primary mechanism at
low metallicities. If O/Fe is constant then the transition between 
primary and secondary production processes would occur at [Fe/H]
$\approx -1$. As such, our quantitative conclusions can not be
definitive. The derivation of the oxygen abundance from the observed
spectra is delicate, specifically at low metallicity. Moreover, non local
thermodynamic equilibrium (NLTE) effects on Fe could affect substantially
the BeB-Fe relationship. We also point out that since magnesium is not
plagued by the same difficulties regarding the stellar yields of Fe, it
would be worthwhile to examine the relationship of the BeB elements with
Mg. New observations are eagerly awaited.

\section{Data Sets}

The BeB-OFe abundance trends encode the history of spallation
in the Galaxy.  Our ability to infer this history 
is completely determined by the accuracy of the abundance
data studied.  It is thus necessary to obtain
high quality data.  Just as importantly, the abundances
must be combined in a systematic and consistent way.

The determination of abundances from
raw stellar spectra requires stellar
atmosphere models.
The atmospheric models require key input parameters, notably
the effective temperature $T_{\rm eff}$ and surface gravity $g$,
and assumptions regarding, 
e.g., the applicability of local thermodynamic equilibrium (LTE).
Unfortunately, there is no
standard set of stellar parameters for the halo stars of interest. 
In practice, different groups derive abundances via
different procedures, which
give similar results but retain systematic differences.
The systematic differences in the data can in fact obscure the  BeB-OFe
trends one seeks. Thus, to derive meaningful BeB fits,
one must systematically and consistently
present abundances derived under the same assumptions
and parameters for stellar atmospheres.

Below, we will present results based for the available BeBOFe data based
on two methods of analysis. We will refer to these as the Balmer line
data and the IRFM data.  The Balmer line data is taken from the work of Axer,
Fuhrmann, \& Gehren \pcite{afg} and Fuhrmann, Axer, \& Gehren \pcite{fag}. 
These authors have studied over 100 dwarf and subgiant stars to determine
a consistent set of stellar parameters which includes the effective
temperature, surface gravity and iron abundance. The effective
temperatures are determined from the synthesis of the first four Balmer
lines. This amounts to a spectroscopic rather than a photometric
determination of the effective temperature. The remaining parameters were
determined simultaneously from the analysis of a number of iron lines.
The temperatures determined this way are typically 100--200 K higher
than many of the photometric determinations in the literature. The
[Fe/H] abundance determined by Axer, Fuhrmann, \& Gehren \pcite{afg} 
are also typically higher than others sometimes by as much as 0.5 dex.
It goes without saying that such differences can have an enormous effect
in trying to establish a trend for BeB vs. Fe.

An alternative method to obtain effective temperatures is achieved by
comparing the bolometric flux and the IR flux at a fixed wavelength.
This method, knows as the IRFM, requires only the
theoretical prediction of the continuum IR flux. There is a slight
dependence on the assumed surface gravity and metallicity.  A large
compilation of nearly 500 stars for which the IRFM was used to determine
temperatures was presented in Alonso, Arribas, \& Martinez-Roger
\pcite{aam1}. Bonifacio and Molaro used these temperatures 
\pcite{bm} to accurately determine Li abundances in 41 plateau stars as a
follow-up of their previous work using Balmer line temperatures (Molaro,
Primas, \&  Bonifacio \cite{mpb}). For Li similar results were found for
the plateau abundance in each of the two methods. In Alonso, Arribas, \&
Martinez-Roger \pcite{aam2}, calibrated expressions for the effective
temperatures were derived based on the earlier tabulation.
Unfortunately, these calibrations do not always lead to the same
temperatures found in the tabulations and differences by as much as 200
K occur. We will present results based on both (1) 
the IRFM temperatures in
the compilation Alonso, Arribas, \& Martinez-Roger
\pcite{aam1}, and (2) based on the calibration. 

We can not overly stress the importance of reliable stellar data.
The Balmer-line data appear to be self-consistent, and are probably the
most reliable.  However, because we will present results based on the
IRFM temperature scales, we would like to point out that there are
significant differences in the data.  To illustrate the point we
take for example the case of the star BD $3^\circ$ 740.
{}From Axer \etal \pcite{afg}, we find this star to have 
($T_{\rm eff}, \ln g$, [Fe/H]) = (6264, 3.72, -2.36).
The beryllium and oxygen abundances for this star was reported by Boesgaard
\etal \pcite{bbe} and \pcite{boes}.  When adjusted for
these stellar parameters, we find [Be/H] = -13.36, and [O/H] = -1.74.
In contrast, the stellar parameters from Alonso \etal \pcite{aam1}
are (6110,3.73,-2.01)\footnote{The surface gravity was recalculated by
Bonifacio \& Molaro \pcite{bm}} with corresponding Be and O abundances of
-13.44 and -2.05.  Garcia-Lopez \etal \pcite{gletal} use a calibrated
IRFM based on Alonso \etal \pcite{aam2} and take (6295,4.00, -3.00).
For these choices, we have [Be/H] = -13.24 and [O/H] = -1.90.
Notice the extremely large range in assumed metallicities and the
difference in the two so-called IRFM temperatures.  While this may not be
a typical example of the difference in stellar parameters,  it is
differences such as this (and this star is not unique) that accounts for
the difference in our results and the implications we must draw from them.

We also note that in addition to the large systematic differences in the
Fe abundances (an order of magnitude in the above example), it has
recently been argued that NLTE effects for Fe, particularly in low
metallicity stars are non-negligible (Th\'evenin \& Idiart \cite{TI}).
At the low end of the metallicity range considered here, they argue for 
a $\sim$ 0.3 dex upward correction in [Fe/H]. We have not included this
correction in the data discussed below. 
  
In the next section we will fit the BeB-OFe abundances over Pop II
metallicities. 
The fits span a metallicity range
$-3 \la \feh \la -0.8$, and 
$-2 \la \oh \la -0.5$ depending on the particular choice of stellar
parameters.  The upper bounds roughly mark
the disk-halo transition,
and the low metallicity bounds 
are just set by the availability of BeB data.

As we indicated, we will consider three sets of stellar parameters to be
used in the BeB vs OFe analysis.  The three are: stellar parameters based
on the Balmer line data of Axer \etal \pcite{afg} -- to be denoted as
Balmer; the IRFM, with data from Alonso \etal \pcite{aam1} -- denoted as
IRFM1; the IRFM with stellar parameters as determined from the
calibrations in Alonso \etal \pcite{aam2} as reported by Garcia-Lopez
\etal \pcite{gletal} and Israelian \etal \pcite{igr} -- denoted as IRFM2. Note
 that at low metallicities this calibration is based on an analytic
formula which is divergent. Thus, the points at low metallicity could be
questionable. Be abundances used in the present work are from Rebolo \etal
\pcite{R}, Ryan \etal
\pcite{r1,r2}, Gilmore \etal \pcite{gil}, Rebolo \etal
\pcite{rebnew}, Boesgaard \& King
\pcite{bk}, Primas \pcite{pth}, Hobbs
\& Thorburn \pcite{ht}, Molaro
\etal \pcite{mbe}, Boesgaard \etal \pcite{bbe}. For each star, and 
for each choice of Balmer, IRFM1,2, we first adjust the Be abundance to a
common set of stellar parameters. Then when multiple observations for a
given star are available these abundances are weight averaged. LTE is
commonly used, since non-LTE (NLTE) corrections for Be are not expected to be
significant. In total, we have data on Be for 37 stars. Not all of the
stellar parameters have been determined in a uniform way for the entire
set of 37 stars.  In the case of the Balmer set, we have parameters from 
Axer \etal \pcite{afg} for 22 stars. This set can be enlarged somewhat
using the data of Boesgaard \etal \pcite{bbe} who have presented results
based on two scales, the King \pcite{King} scale which they claim is
similar to Balmer and Carney \pcite{Carney} which they claim is similar to
IRFM.  Thus for Balmer + King we have data for 30 stars.  
In the case of IRFM1, we also have data on 22 stars. IRFM 1 + Carney
contains 25 stars. Finally for IRFM2, there are only 18 stars, but IRFM2 +
Carney gets us up to 24 stars.  

B abundances are taken from Duncan \etal \pcite{dbor} and Garcia-Lopez
\etal \pcite{gletal} and Primas \etal (1999). NLTE corrections for B
 are significant (Kiselman
\cite{kiss1}, Kiselman \& Carlsson \cite{kiss2}), and are not uniform over
metallicity, giving larger enhancements at the lower metallicities. Thus,
NLTE correction have the systematic effect of flattening the B-OFe
trends. We apply the same procedure as described above for the boron
data.  There are a total of  15 low metallicity stars with boron
measured, of which we have Balmer data for 11 (and Balmer + King on 
13). 9 stars are known for IRFM1 (and 10 for IRFM1 + Carney), and 11
for IRFM2.

\begin{table}[htb]
\caption{Slopes for O/Fe versus Fe.}
\begin{tabular}{ccc}
\hline\hline
 number & method &  slope \\
\hline
22 & Balmer &      $-0.48 \pm 0.16$ \\
31 & Balmer &      $-0.51 \pm 0.10$ \\
36 & Balmer &      $-0.47 \pm 0.09$ \\
22 & IRFM1 &      $-0.45 \pm 0.14$ \\
27 & IRFM1 &      $-0.44 \pm 0.10$ \\
36 & IRFM1 &      $-0.43 \pm 0.09$ \\
22 & IRFM2 &      $-0.35 \pm 0.10$ \\
29 & IRFM2 &      $-0.36 \pm 0.09$ \\
36 & IRFM2 &      $-0.32 \pm 0.08$ \\
\hline\hline
\end{tabular}
\label{tab:O/Fe}
\end{table}

Pop II oxygen abundances, and O/Fe, are critical
for all studies of nucleosynthesis.
The long discussion of O abundances in the literature
illustrates that the path from stellar spectra to
abundances is not a trivial one.  
Three different lines have been used to obtain the 
oxygen abundances with varying results.
These are:
(1) the allowed \ion{O}{1} triplet at 7774 \AA;
(2) the forbidden [\ion{O}{1}] line at 6300 \AA; and
(3) molecular OH lines at 3085 \AA.
The third method was suggested by Bessell, Hughes \& Cottrell \pcite{bhc}
and recently high resolution UV spectra were obtained for many of the
stars with Be and B observations (Nissen \etal \cite{ngeg}, Israelian,
Garc\'{\i}a-L\'{o}pez, \& Rebolo
\cite{igr}, Boesgaard \etal \cite{boes}).
These new data suggest a significant nonzero [O/Fe] trend (versus \feh)
in Pop II stars. 
We use all of the recent data on OH lines and combine the data in a
similar manner as that described for Be. In the two most recent works, a
strong case is made for a varying O/Fe at low metallicity.  The results
of these two groups are in excellent agreement. It should be noted that
the forbidden line continues to show lower O/H abundances as claimed in
the recent work of Fulbright \& Kraft \pcite{ful}.

For O/H, we have a total of 36 stars with low metallicity OH data. There
are 22 stars with Balmer data, and 31 with Balmer + King. There are 22
stars with IRFM1 data and 27 with IRFM1 + Carney. There are also 22 IRFM2
O/H data points and 29 IRFM2 + Carney.
We will define the logarithmic slope $\dofe$ for O/Fe vs Fe by
\beq
\label{eq:O/Fe}
[{\rm O/Fe}] = \dofe \feh + const
\eeq 
Our results for the slopes are given in Table \ref{tab:O/Fe}.
Within the uncertainties these slopes are all consistent with one another.
The IRFM2 slopes are somewhat smaller and in very good agreement with
those found by Israelian \etal \pcite{igr} as should be expected since the
IRFM2 temperature scales are taken from that work.

\vskip 1in

\section{Statistical Tests of BeB Origin}

We will first derive the overall fits for BeB vs both O and Fe using the
data discussed above. 
The Be and B data
can be fit versus \feh\ or [O/H], using logarithmic abundances so that for
example,
\beq
\label{eq:slope_def}
[{\rm Be}] = \dbefe \, \feh \ + \ const
\eeq
We will focus on the logarithmic slopes such as  $\dbefe$;
the same procedure, applied to boron, gives $\dbfe$, etc.
[Be] is defined as $\log$(Be/H) + 12. 
Our results are summarized in tables 
below. 

\begin{table}[htb]
\caption{Slopes for Be versus Fe and O.}
\begin{tabular}{ccccccc}
\hline\hline
  method & number & tracer & slope &  number & tracer & slope\\
\hline
 Balmer & 22 &  Fe &  $1.39 \pm 0.16$ & 19 &  O  &  $1.78 \pm 0.19$  \\
Balmer & 30 &  Fe &  $1.26 \pm 0.11$ & 27 &  O  &  $1.79 \pm 0.16$\\
Balmer & 37 &  Fe &  $1.27 \pm 0.10$ & 31 &  O  &  $1.70 \pm 0.17$\\
IRFM1 & 22 &   Fe &   $1.23 \pm 0.14$ & 21 &   O  &   $1.83 \pm 0.19$\\
IRFM1 & 25 &   Fe &   $1.19 \pm 0.11$ & 24 &   O  &   $1.80 \pm 0.17$ \\
IRFM1 & 37 &   Fe &   $1.21 \pm 0.10$ & 31 &   O  &   $1.49 \pm 0.14$\\
IRFM2 & 18 &   Fe &   $1.18 \pm 0.11$ & 18 &   O  &   $1.36 \pm 0.09$\\
IRFM2 & 24 &   Fe &   $1.15 \pm 0.10$ & 24 &   O  &   $1.35 \pm 0.08$\\
IRFM2 & 37 &   Fe &   $1.18 \pm 0.09$ & 31 &   O  &   $1.29 \pm 0.08$\\
\hline\hline
\end{tabular}
\label{bedata}
\end{table}

\begin{table}[htb]
\caption{Slopes for B versus Fe and O.}
\begin{tabular}{ccccccc}
\hline\hline
method & number & tracer & slope &  number & tracer & slope \\
\hline
Balmer & 11 &  Fe &  $0.78 \pm 0.22$ & 10 &  O  &  $1.23 \pm 0.32$\\
Balmer & 13 &  Fe &  $0.56 \pm 0.14$ & 12 &  O  &  $1.22 \pm 0.28$\\
Balmer & 15 &  Fe &  $0.60 \pm 0.14$ &  &    &  \\
IRFM1  & 9 &   Fe &   $0.73 \pm 0.19$ & 9 &   O  &   $0.98 \pm 0.28$\\
IRFM1 & 10 &   Fe &   $0.63 \pm 0.13$ & 10 &   O  &   $0.94 \pm 0.23$\\
IRFM1 & 15 &   Fe &   $0.70 \pm 0.10$ & 12 &   O  &   $1.00 \pm 0.20$\\
IRFM2 & 11 &   Fe &   $0.72 \pm 0.14$ & 11 &   O  &   $1.02 \pm 0.16$\\
IRFM2 & 15 &   Fe &   $0.70 \pm 0.11$ & 12 &   O  &   $1.00 \pm 0.16$\\
\hline\hline
\end{tabular}
\label{bdata}
\end{table}

\begin{table}[htb]
\caption{Slopes for B/Be versus Fe and O.}
\begin{tabular}{ccccccc}
\hline\hline
method & number & tracer & slope &  number & tracer & slope\\
\hline
Balmer & 9 &  Fe &  $-0.52 \pm 0.24$ & 9 &  O  &  $-0.68 \pm 0.37$\\
Balmer & 11 &  Fe &  $-0.57 \pm 0.16$ & 11 &  O  &  $-0.91 \pm 0.30$\\
IRFM1  & 9 &   Fe &   $-0.49 \pm 0.18$ & 9 &   O  &   $-0.81 \pm 0.27$\\
IRFM1 & 10 &   Fe &   $-0.59 \pm 0.15$ & 10 &   O  &   $-0.96 \pm 0.25$\\
IRFM1 & 11 &   Fe &   $-0.60 \pm 0.15$ & 11 &   O  &   $-0.71 \pm 0.20$\\
IRFM2 & 10 &   Fe &   $-0.57 \pm 0.14$ & 10 &   O  &   $-0.53 \pm 0.21$\\
IRFM2 & 11 &   Fe &   $-0.56 \pm 0.14$ & 11 &   O  &   $-0.49 \pm 0.20$\\
\hline\hline
\end{tabular}
\label{b/bedata}
\end{table}

Note that there is a sharp difference in slopes of BeB vs Fe/H as compared
with O/H. This is implied by the variability in O/Fe at low metallicity.
Also note that for the cases of Balmer and IRFM1, the scaling of Be with
respect to O/H is almost purely secondary.  Boron, on the other hand shows
a strong primary component, which in this case, we would expect to be due
to the $\nu$-process (Woosley \etal \cite{woo1}, Olive \etal \cite{opsv}).
The IRFM2 data still indicates the need for an early primary Be component
as we explain below.

We would also like to call attention to the slopes for B/Be with respect
to either Fe/H or O/H. They are all non-zero (even when the relatively
large uncertainties are taken into account), though there is a
considerable dispersion in the data. This is a departure from the standard
picture of purely secondary GCR nucleosynthesis as well more recent
primary models of BeB nucleosynthesis. This behavior is due in part to
the new Be observations of Boesgaard \etal \pcite{bbe} for some
relatively low metalicity stars. There are now a few cases for which the
B/Be ratio in fact exceed 100 as can be seen in the figures. This is true
for all of the choices of stellar parameters considered. This in fact is a
prediction of the secondary model with $\nu$-process B production (Olive
\etal \cite{opsv}; Fields, Olive, \& Schramm \cite{fos2}).

We are now in position to test the BeB data 
for the presence of two components,  
primary and secondary.
Because physically, BeB production is tied to elements such as CNO, for
the data available, we analyze
only the Be and B trends versus oxygen. 
For $A \in {\rm BeB}$, the
data are fit to 
\beqar
\label{eq:fit_form}
\frac{A}{\rm H} & = & a_1 \, \frac{\rm O}{\rm H} 
	+ a_2 \, \left( \frac{\rm O}{\rm H} \right)^2 \\
  & = & \left( \frac{A}{\rm H} \right)_\odot
        \left[ \alpha_1 \frac{\rm O/H}{({\rm O/H})_\odot}
        + \alpha_2 \left(\frac{\rm O/H}{({\rm O/H})_\odot}\right)^2 \right]
\eeqar
where the fit parameters $a_i$ are expressed in
``scaled'' units $\alpha_1 = ({\rm O}/A)_\odot a_1$ and 
$\alpha_2 = ({\rm O/H})_\odot^2/(A/{\rm H})_\odot \, a_2$.
Note that in eq.\ \pref{eq:fit_form}, the
abundances are {\em not} logarithmic,
but are the (linear)  ratios with respect to hydrogen.
The fit parameters $a_1$ and $a_2$ (equivalently,
$\alpha_1$ and $\alpha_2$) quantify respectively the 
primary and secondary contributions to $A$.
These are given in Table \ref{tab:break}.

\begin{table}[htb]
\caption{Break points for Be versus O.}
\begin{tabular}{ccccc}
\hline\hline
 number & method & $\alpha_1$ & $\alpha_2$ & $\oheq$ \\
\hline
19 & Balmer &  $0.042 \pm 0.003$ &  $2.30 \pm 0.70$ & -1.75 \\
27 & Balmer &  $0.027 \pm 0.027$ &  $2.21 \pm 0.49$ & -1.94\\
31 & Balmer &  $0.027 \pm 0.027$ &  $2.28 \pm 0.52$ & -1.94 \\
21 & IRFM1 &  $0.034 \pm 0.034$ &  $2.14 \pm 0.53$ & -1.79 \\
24 & IRFM1 &  $0.031 \pm 0.027$ &  $2.28 \pm 0.51$ & -1.88 \\
31 & IRFM1 &  $0.050 \pm 0.031$ &  $2.14 \pm 0.54$ & -1.62 \\
18 & IRFM2 &  $0.111 \pm 0.031$ &  $2.57 \pm 0.76$ & -1.37 \\
24 & IRFM2 &  $0.115 \pm 0.031$ &  $2.61 \pm 0.68$ & -1.36 \\
31 & IRFM2 &  $0.122 \pm 0.031$ &  $2.79 \pm 0.70$ & -1.35 \\
\hline\hline
\end{tabular}
\label{tab:break}
\end{table}

The terms in eq.\ (\ref{eq:fit_form}) are equal 
at the metallicity
\beq
\left( \frac{\rm O}{\rm H} \right)_{\rm eq} = \frac{a_1}{a_2} 
  = \frac{\alpha_1}{\alpha_2} \ \left(\frac{\rm O}{\rm H}\right)_\odot
\eeq
with the primary term dominating at 
${\rm O/H} < ({\rm O/H})_{\rm eq}$, and
the secondary term dominating at 
${\rm O/H} > ({\rm O/H})_{\rm eq}$.
Physically, $({\rm O/H})_{\rm eq}$
identifies the epoch at which 
primary cosmic ray sources are overcome by
secondary sources;
we thus refer to this as the ``break point'' 
of the $A$-O trend.  These points (in [O/H])
appear as \oheq\ in Table \ref{tab:break},
and indicate the metallicity at which
the transition from primary to secondary should
occur.  Recall that any primary component present will be dominant at
sufficiently low [O/H].  Therefore we can use the information in the table
to compare the relevant strengths of the primary and secondary
components and constrain models for cosmic ray origin
and acceleration.  As one can see, for both Balmer and
IRFM1, the break point occurs at very low [O/H], in fact at the edge of
data as can be seen from the figures where the data is plotted. 

There is also the question regarding how meaningful these two-parameter
fits are.  To give a quantitative answer to this question requires
an examination of the $\chi^2$ of the various fits in Table
\ref{tab:break}.  In Table \ref{tab:break2}, 
we show the relevant $\chi^2$'s for
the break point fits together with those for fits where either $\alpha_1$
or $\alpha_2$ are forced to be 0. That is the $\chi^2$ for either a purely
linear or purely quadratic evolution.  

\begin{table}[htb]
\caption{$\chi^2$ for the break points in Table 5.}
\begin{tabular}{ccccc}
\hline\hline
 number & method & $\chi^2$ & $\chi^2(\alpha_1=0)~(P_F)$ &
$\chi^2(\alpha_2=0)~(P_F)$\\
\hline
19 & Balmer &  18.4 &  20.2 (78\%) & 30.9 (99.6\%) \\
27 & Balmer &  27.2 &  28.3 (68\%) & 54.1 (99.99\%)\\
31 & Balmer &  34.1 &  35.3 (68\%) & 61.2 (99.99\%) \\
21 & IRFM1 &  13.8 &  15.0 (68\%) & 30.7 (99.99\%) \\
24  & IRFM1 &   18.1 &  19.4 (78\%) & 39.6 (99.99\%) \\
31  & IRFM1 &   34.4 &  39.1 (94\%) & 53.1 (99.96\%) \\
18 & IRFM2 &   24.2 &  56.4 (99.97\%) & 46.6 (99.8\%) \\
24 & IRFM2 &   29.5 &  61.8 (99.99\%) & 54.0 (99.97\%) \\
31 & IRFM2 &  40.0 &  72.7 (99.99\%) & 65.7 (99.98\%) \\
\hline\hline
\end{tabular}
\label{tab:break2}
\end{table}

In order to justify an additional parameter, we must find a significant
drop in the $\chi^2$. The degree of significance can be quantified by the
F-test, which produces a likelihood that the 2-parameter fit is a better
description of the data than the 1-parameter (primary or secondary) fit.
This quantity is also shown in the Table \ref{tab:break2} as $P_F$. For
example, in the first line of that table, for the Balmer case with 19
points, we see that while the 2-parameter fit is clearly superior (with a
probability of 99.6\%) to a straight primary (linear) fit, it is only an
improvement at the 78\% level over a purely secondary fit. That is, in
the latter case, it should be questioned whether or not the 2-parameter
fit is actually better than a purely secondary (quadratic) fit. This
pattern is similar for the IRFM1 data set as well.  In contrast, for the
IRFM2 data, the 2-parameter fit is clearly preferred to either the purely
primary or secondary fits. 

Before concluding this section, we note that use of Fe abundances to
study the BeB evolution presents us with several difficulties. In
addition to the overall uncertainties in the observational determination
of [Fe/H] (with or without NLTE effects), we will also find it 
difficult to model the recently observed O/Fe slopes.  In section 4.4,
we note that most calculated yields predict an overabundance of Fe with
respect to O, particularly for large stellar masses, when compared with
the recent observations. There is however, a substantial uncertainty in
the ejected yield of Fe relative to the mass of Fe trapped in the stellar
remnant. For this reason, it has been suggested (Shigeyama \& Tsujimoto
\cite{ST}) that the use of Mg instead of Fe should be a reliable tracer
for chemical evolution studies.  Indeed, there have been some recent
preliminary observational attempts (Fuhrmann \etal \cite{fuhretal},
Fuhrmann \cite{fuhr}) to obtain Mg data for the halo stars of interest.
 In this context, it happens that Mg is less affected by 
 spectroscopic uncertainties than O; as such Mg might be
a better metallicity tracer than either O and Fe. 
\section{Models}

Using the results of the BeB fits, we can now 
test different scenarios for their non-thermal, spallative nucleosynthesis.
To compute the BeB-OFe trend expected for each scenario,
one must combine models for
cosmic ray/accelerated particle origin, particle propagation, and
chemical evolution.  We describe these in turn.  

\subsection{Cosmic Ray Sources}

We focus on two suggestions for the
origin of accelerated particles:
\begin{enumerate}
\item Standard GCR, which leads to secondary Be and B trends (proportional to
O$^2$).
\item Superbubble accelerated particles (SAP).
The metal-enriched composition of
the superbubbles is reflected in the particles and leads to
primary Be and B production (proportional to O).
\end{enumerate}
For each of these, we must specify the source spectrum 
$Q(E) \equiv dN/dE$ of the accelerated particles,
their composition, and the dependence of the composition
on that of the ambient medium. 
For further discussion see Vangioni-Flam, Cass\'e, \& Audouze
\pcite{vca}.

\subsubsection{Standard Galactic Cosmic Rays}

For many years it has been a common perception that supernova remnants 
are the principal, perhaps predominant, agents
which accelerate galactic cosmic rays up to an
energy of around $10^{15}$ eV. 
Supernova remnants have the energy necessary to satisfy the
cosmic ray energetics requirement if their acceleration efficiency 
is about 0.1;
the current estimate of supernova rates in
our Galaxy can adequately supply the observed cosmic ray density. Such cosmic
rays are presumed to be energized by diffuse shock (Fermi) acceleration, the
shock itself being induced by the impact of the supernova ejecta on the
surrounding interstellar medium (e.g., Ellison, Drury, \& Meyer \cite{edm}). 

We follow here the standard approach, 
which assumes that supernova
blast waves create strong shocks which
accelerate ISM material, with
a composition that retains certain biases.
Specifically, the model follows very closely that of
Fields \& Olive \pcite{fo99a}, but is also similar to 
that of Lemoine \etal \pcite{lem}.
Since the acceleration engines are supernovae,
the source intensity scales as the supernova 
rate and thus (for all but the earlies times) the star formation
rate:  $Q(t) \propto \dot{N}_{\rm II}(t) \sim \psi(t)$. Note that
in the early Galaxy, Type II supernovae dominate.

The source spectrum is that of particles accelerated by strong shocks.
To a good approximation (e.g., Ellison, Drury, \& Meyer \cite{edm}), 
this gives a power law in momentum,
of the form $dN/dp \propto p^{-2}$.  
In energy space, we have $Q(E) = dN/dp \ dp/dE$,
which goes to $q(E) \sim E^{-1.5}$ at low energies,
and $q(E) \sim E^{-2}$ at high energies.
Of course, the full, relativistic $p(E)$ expression is used
for all energies
in the numerical work.
The proton source spectrum appears in 
Figure \ref{fig:spectra}a. This propagated spectrum (see Figure 1b) is
consistent with the one 
 determined on the basis of a refined propagation model fitting
 a large number of observations by Strong \& Moskalenko \pcite{strong}.

The accelerated particle composition in
species $i$ reflects that of the ISM:
\beq
y_i^{\rm CR}(t) \equiv Q_i(t)/Q_p(t) \propto X_i(t)
\eeq
with the following scaling.
The present-day observed cosmic-ray source composition,
$y_i^{\rm CR}(t_0)$,
differs from a solar (and ISM) composition 
in that refractory and/or low first ionization potential elements are
enhanced with respect to the volatile and/or high ionization potential ones
(Meyer, Ellison, \& Drury \cite{med}, Cass\'e and Goret \cite{caca1}, 
 Vangioni-Flam and Cass\'e \cite{vanvan} ).
For example, oxygen has
a cosmic ray abundance 
$y_{\rm O}^{\rm CR,0} = 3.5 \times 10^{-3} = 4.1 {\rm O/H}_\odot$
which is enhanced over that of the ISM,
while helium is relatively
depleted:  $y_{\rm \alpha}^{\rm CR,0} = 0.067 = 0.69 {\rm He/H}_\odot$.
Thus we use the present-day cosmic-ray abundances 
in the scaling rule for the source particles:
\beq
y_i^{\rm CR}(t) = \frac{X_i(t)}{X_{i,\odot}} \ y_i^{\rm CR}(t_0)
\eeq
This scaling has the desired properties that
$y_i(t) \propto X_i(t)$, and
that $y_i^{\rm CR}(t) = y_i^{\rm CR}(t_0)$ when
$X_i(t)= X_{i,\odot}$.

\subsubsection{Superbubble Accelerated Particles}

An accelerated particle component in addition to the GCRs
has been proposed utilizing i) individual
massive exploding stars (Cass\'e \etal  1995) and ii) 
collective explosions in 
superbubbles.
These are large regions of rarefied, ionized, metal-rich
gas associated with star formation regions.
Observationally, superbubbles are identified by their
dense shells of
neutral gas.  The regions interior to these shells are
devoid of neutral gas,
but sometimes show evidence of ionized gas.
Superbubbles probably play a
fundamental role in the structure and energetics of
star-forming regions, and of the ISM in general;
for a review see, e.g., Spitzer \pcite{spitz}, 
Tenorio-Tagle \& Bodenheimer \pcite{ttb}, MacLow and Mc Cray
 \pcite{macmac}, Walker \etal \pcite{marche}.

The structure and dynamics of superbubbles make them favorable
sites for particle acceleration.
The collective effects of
massive stars
which sweep material out into the dense shells,
leaving behind hot, rarefied interiors.
The copious injection of matter and energy by massive stars
provides the ingredients of
a powerful ion accelerator.
Ejecta from core collapse supernovae
generate recurrent but weak shock waves, and turbulence.  These shocks
act as accelerators within the superbubble environment.
Moreover, these regions provide a large
energy reservoir: they have the
necessary power ($10^{52}$ ergs and higher) 
and size to energize a great
number of low energy particles.

Superbubble accelerated particles (SAP) generated this way
could be a significant 
a Galaxy-wide 
component of non-thermal particles
(Vangioni-Flam \etal \cite{vcfo}, \cite{vroc}).
The SAP's would have a metal-rich composition reflecting
that of the superbubbles.
According to recent calculations by Bykov (1999),
the SAP energy spectrum is not very different from that 
of GCR's (but other spectral shapes are possible). 
The SAP's would inevitably produce BeB over to the long
duration of massive star activity in the parent OB associations
($10^5-10^7$ yr),
and thus 
are likely to be a significant 
BeB source.
The production of BeB in superbubbles,
as well as gamma-rays and X-rays (Vangioni-Flam, Cass\'e, \& Audouze
\cite{vca}) is thus well-motivated.
The existence of these particles can be confirmed by
gamma-ray  line observations that will be performed by the European
INTEGRAL satellite. One could even turn the problem around, and use
constraints on primary BeB  to probe the nonthermal component
and physical conditions in superbubbles.

The study of the particle acceleration in superbubbles is only
in its infancy.
Much remains to be explored, including
the relation between the parent molecular clouds and their stellar
products, the structure and distribution
of the environment and its degree of ionization, 
and the parameters of the shocks.  
All of these will affect the composition
and the energy spectrum of the accelerated particles.

The spectrum, $\phi_{\rm SB,int}$, of energetic particles {\em within} the
rarefied superbubble interior has been studied in
considerable detail (Bykov \cite{bykov2}).
Due to the low densities of these regions,
nuclear interactions are slow and thus BeB production in the interiors
is negligible.  The dominant BeB synthesis instead
occurs as the particles escape from the interiors and
traverse the dense shells.  Thus, in deriving BeB production 
rates, the interior flux, $\phi_{\rm SB,int}$, becomes the {\em source} 
for the particle flux within
the shells surrounding them.  That is, 
$Q_{\rm SB,shell} \propto \phi_{\rm SB,int}$.
For the $\phi_{\rm SB,int}$, we use
a simple analytic form which closely follows 
Bykov's \pcite{bykov2} $\zeta = 0.1$ model.
The source spectrum appears in 
Figure \ref{fig:spectra}a. As one can see, it does
not differ substantially from the GCR source spectrum.

The energetic particle composition 
reflects that of the superbubble interior, and thus
is highly enriched in supernova products.
To model this, we take the particle composition to
be time-independent in the early Galaxy.  We take the superbubble abundances to
be the same as the
ejecta of massive stars with primordial $Z=0$ composition.
Specifically, we set the composition to be that
of a $40 \msol$ star, Woosley \& Weaver's \pcite{ww95}
model U40B.  The particles are thus highly enriched (relative to solar) in
O ($y_{\rm O}^{\rm SB} =  32 {\rm O/H}_\odot$), 
and also enriched in He and C.

\subsection{Cosmic Ray Propagation}

We begin with a transport equation which describes the propagation 
and deceleration of nuclei from their sources
through the ISM.
This equation is a general one, and
can be applied to both
GCR and SAP components.
Once launched in space, accelerated
nuclei are deviated by the magnetic field 
irregularities and lose memory of their origin. 
The particles propagate diffusively throughout the Galaxy.
As they traverse the ISM, they suffer ionization energy losses and 
undergo nuclear interactions.
For the lower energy particles,
ionization losses are important and the distances traversed
are short.
High energy nuclei, however, traverse complicated paths whose
length is much larger than the thickness of the Galaxy.
Each time they reach the 
border of the Galaxy (badly defined) the particles have
a small probability of escape.
The Galaxy in this context is like a ``leaky box,''
and we will describe cosmic ray propagation via the
leaky box model. 
Here we briefly summarize
this model, which is discussed in detail
in the context of BeB nucleosynthesis
by, e.g., Prantzos, Cass\'e, \& Vangioni-Flam \pcite{pcv}
and  Fields, Olive, \& Schramm \pcite{fos}.

The leaky box propagation equation can be expressed as
\beq
\label{eq:prop1}
   \frac{\pd}{\pd t}N(E) = Q(E) - \frac{1}{\tau} N(E) + 
   \frac{\pd}{\pd E} \left[ b(E) \, N(E) \right] 
   \stackrel{\rm{SS}}{=} 0 
\eeq
where the last equality assumes a steady state. 
The source (injection) spectrum $Q(E)$ depend
on the cosmic ray model, as discussed in the preceding sections.
Coulomb interactions with 
ambient electrons
lead to an ionization energy loss rate $b(E)$. 
The escape time, $\tau$, is the harmonic mean of the 
confinement time and the nuclear destruction time.
The energy distribution is modified in the course of the 
propagation since energy losses are energy dependent.

Dividing by the mean density of the ISM one gets
\beq
\label{eq:prop2}
  \frac{\pd}{\pd X}\phi(E) = q(E) - \frac{1}{\Lambda} \phi(E)
  + \frac{\pd}{\pd E} \left[ \omega(E) \, \phi(E) \right] 
 \stackrel{\rm{SS}}{=} 0
\eeq
with $\omega(E) = dE/dX$, in ${\rm MeV}/({\rm g} \, {\rm cm}^{-2})$.
The equilibrium spectrum, $\phi$, is solution of this equation.  
BeB production rates and ratios are obtained by integrating
over this spectrum rather than the source spectrum $q$.
Propagated spectra for GCR and SAP models
appear in Figure \ref{fig:spectra}b.

The production rate of $\ell \in {\rm BeB}$ per unit volume is
\beq
\frac{d}{dt} n_\ell = \sum_{ij} n_i 
  \int \ dE \  \sigma_{ij}^{\ell}(E) \ \phi_j(E) \ S_\ell (E)
\eeq
Here $n_i$ is the ISM number density of species $i$,
$\phi_j(E)$ is the solution to eq.\ \pref{eq:prop2} for particle
species $j$,
and $\sigma_{ij}^{\ell}(E)$ is the cross section for
the spallation reaction $i + j \rightarrow \ell + \cdots$
(Read \& Viola \cite{rv}).
The factor $S_\ell(E) = \exp[ - R_\ell(E)/\Lambda ]$
is the probability that the daughter nucleus $\ell$ is stopped
and thermalized in the ISM before it can escape from the Galaxy.

\subsection{Neutrino Process}
\label{sect:nupro}

The neutrino process 
(Woosley \etal \cite{woo1})
provides a means for direct
(primary) \b11 production in supernovae.
As the large neutrino flux streams from the supernova core,
it traverses the overlying shells of material.  
When the flux passes through the carbon shell,
some of the neutrinos collide inelastically with the
\iso{C}{12} nuclei, and make \b11 spallatively,
by removing a nucleon: $\iso{C}{12} + \nu \rightarrow \b11 + p + \nu^\prime$.
Other inelastic reactions create \b10 and \be9, but these
species do not
survive the subsequent passing of the blast wave.
Thus, the \nupro\ provides an elegant, primary source of \b11, but 
does not make Be.

We adopt the \nupro\ yields of Woosley \& Weaver \pcite{ww95},
who found that \b11 is made at significant levels.
The precise yields, however, depend on the neutrino temperature, and hence are
uncertain. Although the \b11 production is very sensitive to the temperature, 
the relative yields for different mass supernovae
are probably better known.  To allow for this
uncertainty, we follow previous work
(Olive \etal \cite{opsv};
Vangioni-Flam \etal \cite{vcfo};
Fields \& Olive \cite{fo99a}) in
taking the \nupro\ yields to be uncertain by an overall factor.
We thus write $m_{ej,11}(m) = f_\nu m_{ej,11}^{\rm WW}(m)$,
where the yield $m_{ej,11}$ of \b11 is a function of progenitor
mass $m$, and scales with the Woosley \& Weaver yield
$ m_{ej,11}^{\rm WW}(m)$.
The scale factor $f_\nu$ will be set by a fit to the data,
as described below, and the resulting value amounts to a constraint
on the neutrino temperature.

\subsection{Chemical Evolution}

The production of BeB depends on the intensity
and composition of both GCR and SAP,
as well as the ISM composition and gas mass.
These quantities are time dependent, and to follow their
evolution requires a model of Galactic chemical evolution.
The chemical evolution model adopted here is
described in detail in
Fields \& Olive \cite{fo99a}, so we will only summarize 
the key points.

For the star formation rate $\psi(t)$
we adopt $\psi = \lambda M_{\rm gas}$, with
$\lambda = 0.3 \, {\rm Gyr}^{-1}$; our results
are insensitive to the details of $\psi$.
We use a power law initial mass function, $\phi \propto m^{-2.65}$,
with $m \in (0.1\msol,100\msol)$.
For simplicity we use a closed box model, i.e.,
without infall or outflow.  
We account for different stellar lifetimes
as a function of mass, and thus
we do not use the instantaneous recycling approximation.

The elements evolved in the code are
H, He, C, N, O, and Fe, as well as BeB.
The high mass (Type II supernova) 
yields are from Woosley \& Weaver \pcite{ww95}.
The intermediate mass (AGB) yields are from 
van den Hoek \& Groenewegen \pcite{vdhg}.
In both cases, the full metallicity dependence of
the tabulated yields is used.

The different roles of O and Fe deserve mention.
We stress that the oxygen yields are the most critical,
since as we have argued O is the better 
diagnostic for BeB origin, and thus
the BeB-O trends are the main focus 
throughout this paper.
The adopted O yields 
of Woosley \& Weaver \pcite{ww95} are
in broad agreement with the work of 
Thielemann, Nomoto, \& Hashimoto \pcite{tnh}.
Furthermore, the O yields are not strongly altered 
in explosive nucleosynthesis, nor are they sensitive to
the mass cut.  

The yields of iron, on the other hand, 
are much more uncertain.
However, because Fe has been shown in all previous work, we
also will show this for
comparison.
Our treatment of Fe production follows that
of Fields \& Olive \pcite{fo99a}, and is driven by the new O/Fe data,
The observed O/Fe-Fe relation (eq.\ \ref{eq:O/Fe}) is key
to interpreting the BeB-Fe slopes, so
we adopt this relation, and
solve for iron:
\beq
\label{eq:Fe-scale}
{\rm Fe/Fe_\odot} \propto ({\rm O/O_\odot})^{1/(1+\dofe)}
\eeq
Since O yields and evolution are fixed in the model,
this strategy amounts to fixing the effective yield for Fe. 
Note that the non-constancy of O/Fe (i.e., $\dofe \ne 0$)
demands that the derived Fe yield is not proportional to the
O yield.

In other words, we can derive the effective Fe yields needed to 
agree with the O/Fe data and the O yields.
The effective Fe yield $\mejfe$ is the mean Fe mass
ejected per supernova, which is the total Fe ejection rate
divided by the supernova rate:
\beq
\mejfe \equiv \frac{E_{\rm Fe}}{\dot{N}_{\rm II}}
\eeq
The iron mass ejection rate can be inferred as follows.
The usual chemical evolution expressions give
$E_{\rm Fe} = M_g \dot{X}_{\rm Fe} + E {X}_{\rm Fe}$. 
We compute $\dot{X}_{\rm Fe}$ via eq.\ \pref{eq:Fe-scale},
which is imposed via 
$\dot{X}_{\rm Fe}/{X}_{\rm Fe} = (1+\dofe)^{-1}\dot{X}_{\rm O}/{X}_{\rm O}$,
where $\dot{X}_{\rm O}$ is computed as usual from the high-mass yields
within the model.
Thus $\mejfe$ is an effective yield, averaged over the mass function,
at each epoch $t$.

Figure \ref{fig:yields} shows the time evolution of
$\mejfe$, and for comparison, 
a similar effective yield for O.
At any give time, the mass range includes all of
those stars whose mass $m$ is larger than
$m_t$, the mass whose lifetime $\tau(m)=t$.  Thus,
the average
includes an increasingly larger mass range of stars
as time goes on.
We therefore also plot these results as a function of
$m_t$ in the lower panel of Figure \ref{fig:yields}.  
We see that the effective
O yield starts high ($>10 \msol$) when the progenitors
have high mass, and then levels off at
$10 \msol$, where the average then includes
all supernovae.  In contrast, Fe
starts low and then builds up.  Within the
$m > 10 \msol$ regime, the Fe yield is not constant,
but spans a large range, to include values at lower
masses which are of the order of Fe yields estimated
for observed supernovae,
though somewhat lower than the yields of
Woosley \& Weaver \pcite{ww95}. Above 60 \msol, the Fe yield is assumed to be
negligible.  Below $m < 10 \msol$, $\mejfe$ continues to rise
to values approaching $1 \msol$.  This 
rise thus coincides with the epoch when Type Ia supernovae
should begin to contribute to iron enrichment.
The Type Ia iron yields are expected to be
nearly constant per explosion event, and to approach $1\msol$;
thus, one can view the $m < 10 \msol$ behavior of $\mejfe$
as a measure of the increasing
ratio of Type Ia/Type II supernova events.

It is noteworthy that the 
inferred Fe yields are at variance with the 
yields that Shigeyama \& Tsujimoto \pcite{ST}
inferred from 
Mg/Fe ratio in ultra-metal-poor
halo stars (and from the light curves of
supernovae, which however arise from a different
population than the halo stars).  While the trend in 
$\mejfe$ is inferred from O/Fe to be decreasing
with increasing progenitor mass, the trend
inferred from Mg/Fe is found to do just the opposite.
This difference is indeed striking but
not surprising, since Shigeyama \& Tsujimoto \pcite{ST}
adopted a Mg/Fe relation that {\em decreases} towards
low metallicity, 
rather than increases as O/Fe does.
This result discrepancy indicates the need to get good observational
constraints on the O-Fe-Mg relations.

An alternative model has been proposed in which the O/Fe
relationship is obtained by assuming differential mixing of Be, O and Fe, the 
 latter being locked up into grains
 (Ramaty and Lingenfelter 1999, Ramaty \etal  \cite{scul}). With the nominal yields
 of Woosley and Weaver (1995), they obtain an acceptable O/Fe 
 evolution at the expense of an additional free parameter: the mixing time.

\section{Model Results}

In presenting results, we will compare calculated
Be and B trends with observations.
In doing this, we focus on 
oxygen as a metallicity indicator. 
Using oxygen rather than iron
has several theoretical advantages. Oxygen
is a good tracer of the cosmic ray
accelerators (Type II supernovae).
Also, oxygen (and to a lesser extent C and N) 
provides the main raw material from which
Be and B are carved by spallation.
And while both oxygen and iron are both made by
Type II supernovae, the oxygen yields
are considerably less uncertain as they
are dominated by well-understood hydrostatic burning
processes, as opposed to the iron yield which
depends sensitively on the details of the 
(poorly understood) explosion mechanism and mass cut.
However, as noted above, Mg may be a promising 
evolutionary tracer:  from the theoretical point of
view, Mg shares the same robustness in prediced yields
that O does, since it is not affected by the poor
knowledge of the position of the mass cut between the 
neutron star and the supernova ejects on one hand;
from the observational point of view, Mg 
abundances depends less on the details of the stellar atmospheric models
(for instance it is less subject to NLTE effects, Fuhrmann \pcite{fuhr}).

\subsection{Be and B Normalization}

The three components in the BeB evolution lead
to three free parameters:
\begin{enumerate}
\item 
The normalization of the GCR component.  
We scale the GCR flux via 
$\Phi(t) = f_{\rm GCR} [\psi(t)/\psi(t_0)]\Phi_0$, where
the star formation rate scaling, 
$\psi(t)/\psi(t_0)$, is computed self-consistently
in the model, and $\Phi_0$ is the total flux in present-day cosmic rays,
as in Figure 1. 
    
\item The SAP (energetic particle) normalization.
We put $\Phi_{\rm SAP}(t) = f_{\rm SAP} \Phi(t)$.

\item The \nupro\ normalization.  The scaling $f_\nu$ is
described in \S \ref{sect:nupro}.
\end{enumerate}
Each of the $f_i$ are free, in that 
there are independent constraints on the normalizations,
but we do have some independent constraints on their values.
For example, the GCR normalization is essentially fixed
by the present Galactic average GCR flux, which
is constrained by GCR flux measurements in the vicinity of the earth at high energy
($>1$ GeV) as well as by $\gamma$-ray observations (see Strong and Moskalenko 
\pcite{stro2}). 
Thus, the GCR normalization
is not arbitrary, but on the other hand,
since the present-day Galactic average flux is uncertain to within
a factor of a few (Mori \cite{mori}).
We find that
the scaling is $f_{\rm GCR} = 2.1$, well within
the uncertainties of the 
Galactic average GCR flux.  Note that
the other two normalizations are more poorly constrained;
indeed, the BeB data are probably the best way to
constrain them.

We thus need three pieces of data to fix the normalizations;
then the resulting Be-B-O trends are predictions.
This is done as follows
\begin{enumerate}
\item 
A break point \oheq\ is chosen to match that of the data set being
fit. 
This fixes the {\em relative} strength 
$f_{\rm SAP}$ of the GCR vs SAP components.  
Thus, the evolution of the
species that have only GCR and SAP sources, namely
Be and \b10,
are fixed up to an overall scale factor.

\item
The Be evolution is normalized to reach
$({\rm Be/H})_\odot$ at $({\rm O/H})_\odot$,
which fixes $f_{\rm GCR}$.  
Thus Be-O is forced to have the right \oheq,
and must go through the solar point.\footnote{
Fortunately, the solar abundances of Be and B 
are now well established: 
there is now good
agreement between the photospheric and meteoritic abundances of both
beryllium and boron. After a long debate, the situation has finally
been settled: Be and B are undepleted in the solar photosphere.}
We have now fixed the GCR and SAP components completely,
only using the Be break point and solar point.
In doing this,
\b10 is completely determined, and
\b11 is determined up to the \nupro\ contribution;
both of these are {\it predictions} of the model.

\item
Finally, the \nupro\ contribution to \b11 is
scaled so that the predicted
\b11/\b10 ratio agrees with the meteoritic value
at solar metallicity; this fixes $f_\nu$. Now the model is completely
determined, and the scalings with O cannot be adjusted
further.  
\end{enumerate}

Note that this procedure
does {\em not} force elemental B to go through 
the solar point.
As we will see, the models nevertheless {\em do}
pass through or very near the solar point,
which should be viewed as a success of the scenarios being tested.

\subsection{Comparing Theory and Observation}

To compare theory and observation we have run
models following the procedure just described, and
we now plot them with the observations
for the different sets of stellar parameters.

As discussed above, the break point \oheq\
is an input to the model.  However, it
is not an entirely free parameter, as is evident when
applying a two-component (linear plus quadratic)
fit to the Be-O points from the
GCR model alone.  We find that this model
has a small but nonzero linear component,
$\alpha_1 = 0.043$, in addition to the expected large quadratic term,
$\alpha_2= 3.8$ value. 
The small linear component results from the fact,
shown in Fields \& Olive \pcite{fo99a}, that
the Be-O slope is slightly less than 2,
due to small astration effects.  This flattening
leads to a small but nonzero linear component in the fit.
Because $\alpha_1 > 0$, the GCR model {\em alone} has
a break point at $-1.94$, without any addition
of the primary process.  Thus, there is no way to
arrange a break point below $-1.94$, 
since adding a purely primary component only raises the
value of $\alpha_1$ and thus the 
break point, defined by $\alpha_1/\alpha_2$.  

In making the comparison between theory and observations, we have chosen
the break point for each method corresponding to the {\em lowest} number
of data points. 
This data subset uses stellar atmospheric parameters 
which have been derived directly from
the prescription discussed, and therefore
should be most free of systematic effects.   
For example, in the Balmer
case, only 19 of the total 31 stars are truly what we have called Balmer.
Recall an additional 8 were described as Balmer + King in section 2 and may have
additional systematic effects entering which we are trying to avoid.  This
intermediate set (as well as the full data set) were used to compute slope values,
etc., for comparison only. 

The Balmer data have 
$\oheq = -1.75$ as given in (Table \ref{tab:break}); the results
of a BeB model with this \oheq\ value appears
in Figure \ref{fig:BeB-O_balm}.  
As it happens, this break point is close to
that of the GCR component,
and thus there is only a small SAP component
for this model. Note that in this figure and those that follow, the SAP
component and the GCR components cross at value of [O/H] below \oheq. 
This is due to the small primary component in the GCRs themselves 
($\alpha_{1,{\rm GCR}}=0.043)$, as discussed above.
Thus
\oheq~ is determined by the sum of the two primary components, relative to
the secondary component. One can see that the Be-O trend in this model
fits the data quite well. The B-O model contains in addition the
\nupro.  The lower panel of Figure
\ref{fig:BeB-O_balm}. plots the different components as well as the 
evolution of total abundance which sums
the components.  
We see that B-O is dominated by the primary \nupro\
at low metallicity, and by the secondary GCR production at
high metallicity, as expected.  However, even at present
the \nupro\ component is not negligible, being
lower than the total GCR component by about a factor of 2
(as is indeed demanded by the fit to the \b11/\b10 ratio).
We have scaled the \nupro yields of Woosley \etal \pcite{woo1} downward by
a factor of $f_\nu = 0.4$ here and, as it turns out,
 in the other models as well. This scaling
also keeps the production of \li7 small enough so as not to affect the Spite plateau
as shown in Vangioni-Flam \etal \pcite{vcfo}. In Figure 3, and in the figures that
follow, the point shapes indicate which data sample the stars fall into (e.g., pure
Balmer, Balmer + King, or full).

Figure \ref{fig:BeB-O_irfm1}
plots Be and B versus O for IRFM1,
which has a break point $\oheq = -1.79$. 
The upper panel shows Be-O, 
and
is very similar to the Balmer case discussed above
due to the similarity in the break points.
Again, the intrinsic linear portion of the GCR
component influences the break point, which is why
the break point is higher than the metallicity 
at which the GCR and SAP components are equal.
The quality of the Be-O fit is clearly good.
In the case of B-O (lower panel), 
the \nupro\ contributes as well, leading to
a GCR dominance at a higher metallicity than for Be-O.
Unlike Be-O, the B-O fit now seems to be a poor one.
The overall shape is fine, but the curve is clearly high.

Figure \ref{fig:BeB-O_irfm2} plots BeB-O evolution for
IRFM2, which has $\oheq = -1.37$.  In this case, the transition from
SAP dominance at low metallicity to GCR dominance at high metallicity is evident. The
results here are qualitatively similar to the IRFM1 case:
the Be-O evolution is satisfactory, while the
B-O evolution seems high. 

In the next set of figures, we show plots of B/Be versus O.  
At the lowest metallicity in the Balmer case, Be is secondary while
B is dominated by the \nupro\ and is thus primary.
Hence, the B/Be ratio changes as ${\rm B/Be} \propto {\rm O}^{-1}$
and has a 
log slope of -1,
as seen in Figure \ref{fig:BBe-O_balm}. 
Interestingly, even with this rise, the model actually falls short
of the two highest B/Be ratios, though the qualitative behavior
is correct.  
At high metallicity, the GCR component dominates both Be and B,
and the ratio flattens. In the IRFM case, at very low metallicity, 
  both Be and B
are dominated by the SAP component, and again the curve flattens. 
The behavior of B/Be for the other two cases 
are shown in Figure \ref{fig:BBe-O_irfm1} (for IRFM1) and  in
Figure \ref{fig:BBe-O_irfm2} (for IRFM2). In the latter case the break point
is higher and thus there is only a narrow range
in [O/H] where B/Be drops, and so the overall change
in B/Be is smaller.

In the IRFM1 case, B/Be is
also far from the very high B/Be data points
at the lowest metallicities.  This illustrates the
power of the B/Be ratio to constrain the models.
The data is still poor however.  
While the situation is presently too
uncertain to clearly exclude the models
presented, it is clear that more B/Be data at low metallicity
will play a decisive role in determining 
Be and B origins. 

In Table \ref{tab:prod_ratios}, we show the production ratios of GCR and SAP
at solar metallicity.
That these are so close traces back to the
very similar
energy spectra (in spite of their different
composition).

\begin{table}[htb]
\caption {Production ratios for GCR and superbubble accelerated particles (SAP)}
\begin{tabular}{ccc}
\hline\hline
Ratio & Present ($Z=Z_\odot$) GCR &  SAP \\
\hline
    \li6/\be9	&	5	&		5.8 \\
     \b10/\be9    &       5        &              4.9 \\
     \li7/\li6     &    1.4       &               1.2 \\
     \li7/\be9       &    7       &               6.8 \\
     \b11/\be9        &  12         &            11.7 \\
     \b11/\b10      &     2.5       &            2.4  \\
       Li/Be         &     12        &           14    \\
        B/Be          &     17      &             16.6  \\
\hline\hline
\end{tabular}
\label{tab:prod_ratios}
\end{table}

\subsection{Energetics}

As emphasized by Ramaty \etal \pcite{rklr,scul},
a study of the energy budget associated
with BeB production models
can give additional insight.
There are several ways to proceed,
but the key point is to
link the energy in
accelerated particles required to make the BeB
with the available energy budget
in the form of supernovae.
The power of the analysis is in
judging whether the energy requirement
changes with metallicity (and thus epoch);
in particular, it is important to see whether
there is too large an energy requirement
in Population II.

We find that there is not an energetics problem
in the models we consider.
One way to see this is as follows.
Given a model, we wish to compute the
accelerated particle energy input per supernova: 
\beq
\label{eq:E_cr}
\ecr = \frac{\decrt}{\dnsn}
\eeq
where $\decrt(t)$ is the total energy going into
particle acceleration throughout the Galaxy per second, and
$\dnsn$ is the supernova rate.
The direct approach is simply to note that
in both the GCR and primary components,
supernovae are the power source of energetic particles: 
$\decrt \propto Q \propto \dnsn$.
Since the power input is proportional to the supernova
rate, their ratio $\ecr$ is constant, 
independent of metallicity.  The models always require
the same accelerated particle energy per supernova. 
Note that this approach simply examines the internal
consistency of the model.  We have not used the 
properties of the Be data.

We can use the data, however, and get the same conclusion.
One can derive the $\ecr$ scaling
{\em observationally} by relating the Be-O and Be-Fe data to the 
energy budget.  
This comes about by using the definition of $\ecr$ above
(eq.\ \ref{eq:E_cr}) and a ``chain rule'' expansion: 
\beq
\label{eq:Eobs}
\left(\ecr\right)_{\rm obs} =  
   \left(\frac{\decrt}{\dot{M}_{\rm Be}}\right)_{\rm model} \
   \left(\frac{\dot{M}_{\rm Be}}{\dot{M}_{Z}}\right)_{\rm obs} \
   \left(\frac{\dot{M}_{Z}}{\dnsn}\right)_{\rm model}
\eeq
for the metal tracers $Z \in {\rm O,Fe}$.
As we have indicated, the first and last terms must be taken from
the model (note that in Pop II, $\decrt/\dot{M}_{\rm Be} \simeq {W}/Q(Be)$ in
the Ramaty \etal \cite{rklr} notation).  The middle term can be related
to the data, since 
$\dot{M}_{\rm Be}/\dot{M}_{Z} \propto {\rm Be}/{Z}$
as long as Be can be expressed as a power law in $Z$.
All theory terms (both denominators and numerators) are
individually calculable, so we can now rearrange terms:
\begin{eqnarray}
\label{eq:Erearr}
\left(\ecr\right)_{\rm obs} & =  &
   \left(\frac{\decrt}{\dnsn}\right)_{\rm model} \
   \left(\frac{\dot{M}_{\rm Be}}{\dot{M}_{Z}}\right)_{\rm obs} \
   \left(\frac{\dot{M}_{Z}}{\dot{M}_{\rm Be}}\right)_{\rm model} \\
  & \propto & \frac{({\rm Be}/Z)_{\rm obs}}
                   {({\rm Be}/Z)_{\rm model}}
\label{eq:Esimp}
\end{eqnarray}
Equation \pref{eq:Esimp} follows from
eq.\ \pref{eq:Erearr} due to the intrinsic constancy
of $\decrt/\dnsn$ in the models; now, however, something new
has been added in that the observed Be/$Z$ ratio appears explicitly.
In particular, we see that the scaling of $\ecr$ with
metallicity depends on the ratio of observed to predicted 
Be/$Z$.  If the observations are well matched by the predictions,
then the ratio remains constant, as does $\ecr$.   
In other words, the constancy of $\ecr$ is determined by 
the goodness of the Be-$Z$ fit.  Thus, 
{\em all models which fit the data
are able to avoid an energetics problem.}
This applies quite generally, to the 
models presented here as well as others (e.g., Ramaty \etal \cite{scul})
which fit the data.\footnote{Strictly speaking, 
the scaling argument we have given says that
models which do not have an
energetics problem at the {\em present epoch} 
and fit the Be/$Z$ data will also satisfy the
energy requirements in earlier epochs.} 

This argument shows that the models in question
are energetically sound, simply due to their agreement with the data; however, 
a more detailed analysis is instructive.
As we have argued, it is preferable to use the
O data as the metal tracer $Z$, 
as this is the most direct diagnostic of BeB origins. 
As we have emphasized, and shown in Table \ref{bedata},
the observational data clearly show that Be/O is not constant, but rather
$({\rm Be/O})_{\rm obs} \sim {\rm O}^{0.8 \pm 0.2}$, 
where the exact power depends
on the atmospheric model used.  This factor alone
would lead to a {\em decrease} in $\ecr$ at low metallicity.
However, when the GCR is dominant over SAP's, (i.e., for
metallicities $\oh > \oheq$),
the ``production efficiency'' factor scales
as $\decrt/\dot{M}_{\rm Be} \sim \dnsn/({\rm O}\dnsn) \sim {\rm O}^{-1}$.
This reflects the reduced efficiency for GCR Be production 
in the early Galaxy:  the paucity of O targets in the 
early ISM leads to a lower BeB production for each supernova's
complement of accelerated particles; this factor alone would
lead to an {\em increase} in $\ecr.$ 
Thus, the first two
factors in eq.\ \pref{eq:Eobs}
offset each other, which is largely why $\ecr$ is constant.

We must consider several smaller and more subtle effects 
to complete our analysis of eq.\ \pref{eq:Eobs}.
The last factor in eq.\ \pref{eq:Eobs}, the ``average yield'' 
$\dot{M}_{Z}/\dnsn$ has a weak dependence on O, rising
slightly towards low metallicity.  
However, 
we have until now assumed that 
$\dot{M}_{\rm Be}/\dot{M}_{Z} \propto ({\rm Be}/{Z})_{\rm obs}$.
This is a good approximation in Pop II, where
Be evolution is dominated by production only,
i.e., $\dot{M}_{\rm Be} \sim Q({\rm Be})$. 
But  
in Pop I, astration effects become important since many low-mass
stars begin to die and return their astrated, Be-free gas,
which dilutes the ISM Be abundance.  
Astration manifests itself in the Be-O and Be-Fe relations by
the turnover of Be at high metallicity:  Be production is
overtaken by destruction.  
Thus, for Pop I, the change in Be gas mass
is lower than the cosmic ray production term
would give:  $\dot{M}_{\rm Be} < Q({\rm Be})$.
Pop I Be production is, in this respect, less efficient than in Pop II
where astration is small and $\dot{M}_{\rm Be} \simeq Q({\rm Be})$.
Thus, the Be-$Z$ relation is flattened
in Pop I, and the extrapolation of $\ecr$ via eq.\ \pref{eq:Eobs}
back to Pop II overestimates the energy needed 
if this astration is neglected.  Finally, 
at very low metallicity, $\oh \la \oheq$, 
the SAP component takes over the energy budget.
Thus, just as SAP's are primarily responsible for BeB
production below \oheq, they are also the dominant
energy source in this range.

Finally, we consider using $Z = {\rm Fe}$ in 
eq.\ \pref{eq:Eobs}. 
To reiterate, the question regarding the energetic
problem, rests on $\ecr$. 
This quantity may be constant, even though
$W/Q({\rm Be})$ 
is not constant, depending on the observed Be/Fe ratio 
and the ratio of Be/Fe produced.
Since the Be-Fe slopes are larger than
(though close to) 1,
Be/Fe increases with Fe.
More significantly, the Be/Fe production ratio
does not scale as Fe itself, but rather
$\dot{M}_{\rm Be} \propto O$ and
${\rm O} \sim {\rm Fe}^{0.5}$.
Similarly, the iron yield per supernova
may also not be constant if the
yield of Fe falls off at large stellar masses
(see Fig \ref{fig:yields}).  Both of these facts
simply reflect the fact that O/Fe is not constant in
Pop II.  Thus, models in which $W/Q(Be)$ is
not constant, can not a priori be disregarded.

\section{Conclusions} 

The history of the accelerated particles 
within the Galaxy is
coded in the
evolution of spallogenic Be and B.
Recent theoretical studies of particle acceleration
suggest that the Galaxy
may have two components of nonthermal particles.  
These two populations yield different BeB evolution,
GCR particles lead to secondary ($\propto {\rm O}^2$) production,
while nuclei accelerated in superbubbles
give a primary ($\propto {\rm O}$) production.

We have searched for 
the corresponding evidence of two components
in the evolution of Be and B versus O and Fe abundances
in low metallicity
halo stars.
Specifically, we have made
two-component fits of abundance data, 
to identify the presence of primary 
and secondary BeB components,
and to quantify their relative strength.
This analysis has been carried out 
using consistent stellar atmosphere parameters
based on Balmer lines and the infrared flux method (IRFM).  
Some results depend on the method used to get the
atmospheric parameters, while others do not. 
In both Balmer and IRFM methods, the secondary BeB component
is seen to dominate for $\oh \ga -1.4$.  However,
the status of the primary component depends on 
the atmospheric parameters chosen.
When the Balmer treatment is used, 
a primary mechanism is not required over the observed metallicities,
whereas it is required in the IRFM2 case, where it dominates
at $\oh \la -1.4$.

We use the BeB-OFe fits to constrain the 
theories, whose predictions we model by combining
cosmic ray nucleosynthesis and Galactic chemical evolution.
The GCR component has the observed present-day spectrum,
is assumed to be accelerated from supernovae,
and to have a composition which scales with 
that of the ISM.
For the case of superbubbles, the energy spectrum
has been taken from Bykov (1999), and is quite similar to the
that of the GCRs.
Thus there a little hope to differentiate the two
components on an isotopic basis, since the production ratios of
superbubbles and GCR are very similar in spite of their different
composition (C/O = 0.09 against C/O = 0.81 respectively).
The two-component BeB-O fits constrain the models
by demanding that: (1) the GCR production dominates
over the superbubble component
at $\oh > -1.4$ in all cases, and in the Balmer case
it is unclear whether the SAP component is needed;
(2) the secondary, GCR component requires that the
mean GCR flux at present factor of 2.1
higher than current estimates; given the uncertainties,
this is quite plausible; and (3) the \nupro\
is needed to produce \b11, but at a level
that is 40\% of the fiducial Woosley \& Weaver \pcite{ww95}
yields; again, this is well within the uncertainties of
the yield calculations.

Our results can be further tested, and the constraints tightened,
by several types of observations.
Simultaneous and high signal/noise measurements of Be,
B, O, and Fe, are called for. 
These must be interpreted within
detailed and consistent atmospheric models including accurate
 NLTE corrections. The current situation is not definitive, since the 
conclusions regarding the primary component depend sensitively
on the atmospheric parameters adopted.
A critical goal is to determine Be and B 
abundances--as well as key ratios such as B/Be, Li/Be, and \b11/\b10--at 
very low metallicity, where the primary and secondary 
origins differ most in their predictions.  
It is just as important to firmly establish the nature and evolution of O/Fe;
other metallicity indices, such as Mg, could
help to clarify this question. These should be analyzed systematically
as we have tried to do with oxygen.  
Gamma-ray line astronomy should also help to confirm the
possibility of acceleration of freshly synthesized nuclei in
superbubbles and OB associations.  This will be a key objective for
the European INTEGRAL mission. 
Thus, we are pleased to report that
the nature of 
LiBeB evolution, 
is a problem that is ripe for new and precision
observations, which can 
go far to reveal the origin and history of 
accelerated particles in the Galaxy.

\acknowledgments
We thank Andrei Bykov for graciously making available
data files for his published spectra. We thank
Roger Cayrel and Evan Skillman for fruitful discussions.
This work
was supported in part by DoE grant DE-FG02-94ER-40823 at the University of
Minnesota.

\nobreak

\newpage

\centerline{\bf Figure Captions}

\begin{enumerate}

\item
\label{fig:spectra}
Source and propagated spectra for
accelerated particles. \\
{\it Top Panel}:  Source spectra $Q(E)$ for the GCR (solid curve)
and SAP (dashed curve) components.  $E$ is the kinetic energy per
nucleon in MeV/n, while the units of $Q(E)$ are arbitrary,
and the relative scaling of the two components has been
adjusted for clarity. \\
{\it Bottom Panel}: Propagated spectra corresponding to the
sources in the top panel.

\item
\label{fig:yields}
Oxygen (solid line) and iron (dashed line) yield estimates, calculated from the numerical model.
For derivation see discussion in text. \\
{\it Top Panel}:  Yields versus time. \\
{\it Bottom Panel}:  Results from top panel translated to a mass scale
via the stellar mass-lifetime relation $\tau(m)$. 

\item
\label{fig:BeB-O_balm}
Be {\it vs} O ({\it top panel}) and B {\it vs} O ({\it bottom panel}).  
Data shown are the Balmer
points, which are found (Table \ref{tab:break}) to have a break
point as indicated.  Models are adjusted to have the
break point and O/Fe slope of these data.

\item
\label{fig:BeB-O_irfm1}
As in Fig.\ \ref{fig:BeB-O_balm}, for the IRFM1 data
and corresponding model.

\item
\label{fig:BeB-O_irfm2}
As in Fig.\ \ref{fig:BeB-O_balm}, for the IRFM2
data and corresponding model.

\item
\label{fig:BBe-O_balm}
The B/Be ratio {\it vs} O.  Shown are the Balmer data
and the corresponding model.

\item
\label{fig:BBe-O_irfm1}
As in Fig.\ \ref{fig:BBe-O_balm}, for the IRFM1
data and corresponding models.

\item
\label{fig:BBe-O_irfm2}
As in Fig.\ \ref{fig:BBe-O_balm}, for the IRFM2
data and corresponding models.

\item
\label{fig:BeB-Fe_balm}
Be and B vs Fe, for the Balmer data and break point.  
As discussed in the text, Fe in the model is obtained by
scaling from O, 
using the observed relation
$\oh = \dofe \feh$, with $\dofe$ from Table \ref{tab:O/Fe}.

\item
\label{fig:BeB-Fe_irfm1}
As in Fig.\ \ref{fig:BeB-Fe_balm}, for the
IRFM1 data and corresponding models.
 
\item
\label{fig:BeB-Fe_irfm2}
As in Fig.\ \ref{fig:BeB-Fe_balm}, for the
IRFM2 data and corresponding models.

\end{enumerate}

\newpage

\begin{figure}[!htb]
\begin{center}
\leavevmode
\epsfxsize=160mm
\epsfbox{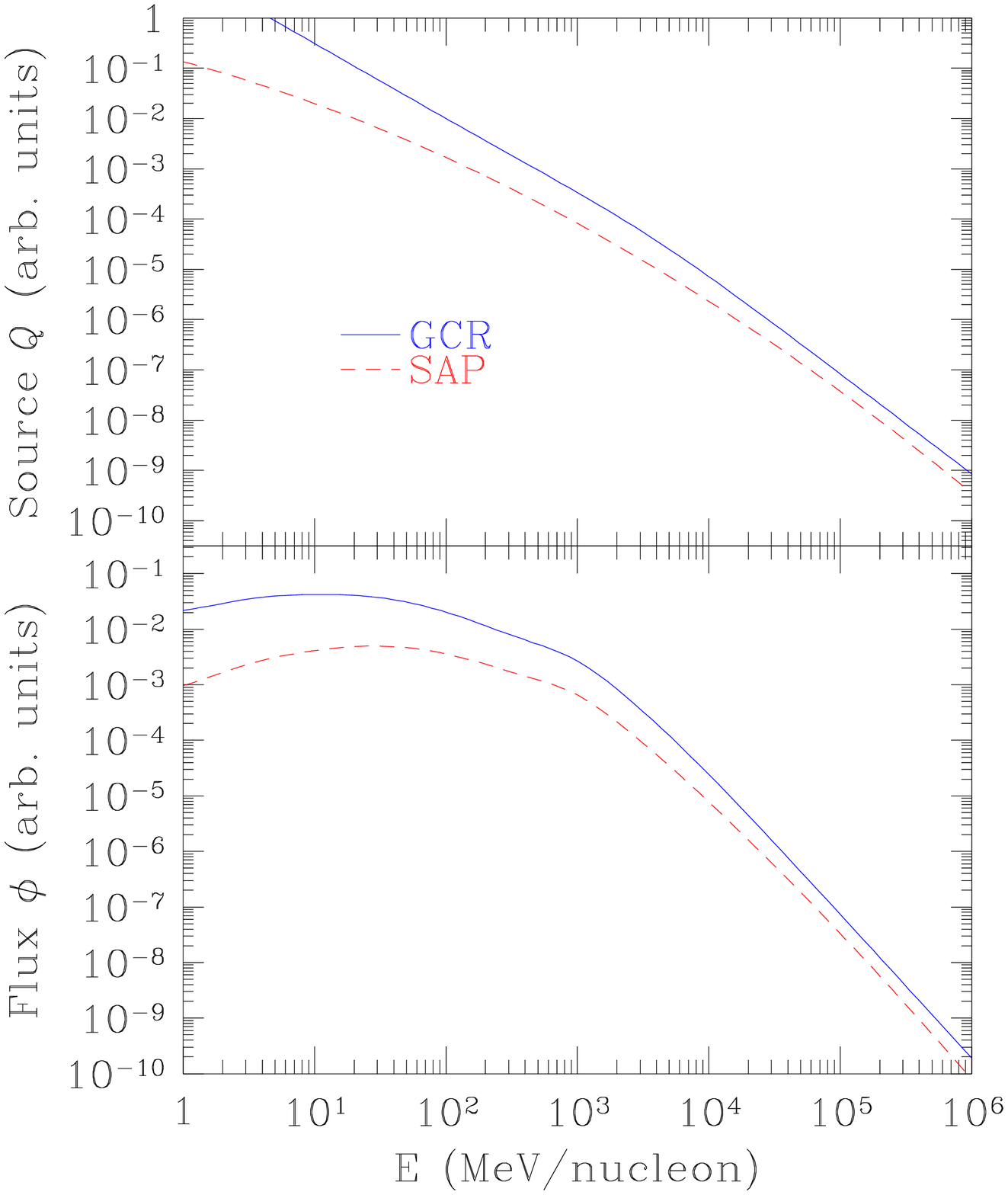}
\end{center}
\caption{}
\end{figure}

\newpage

\begin{figure}[!htb]
\begin{center}
\leavevmode
\epsfxsize=160mm
\epsfbox{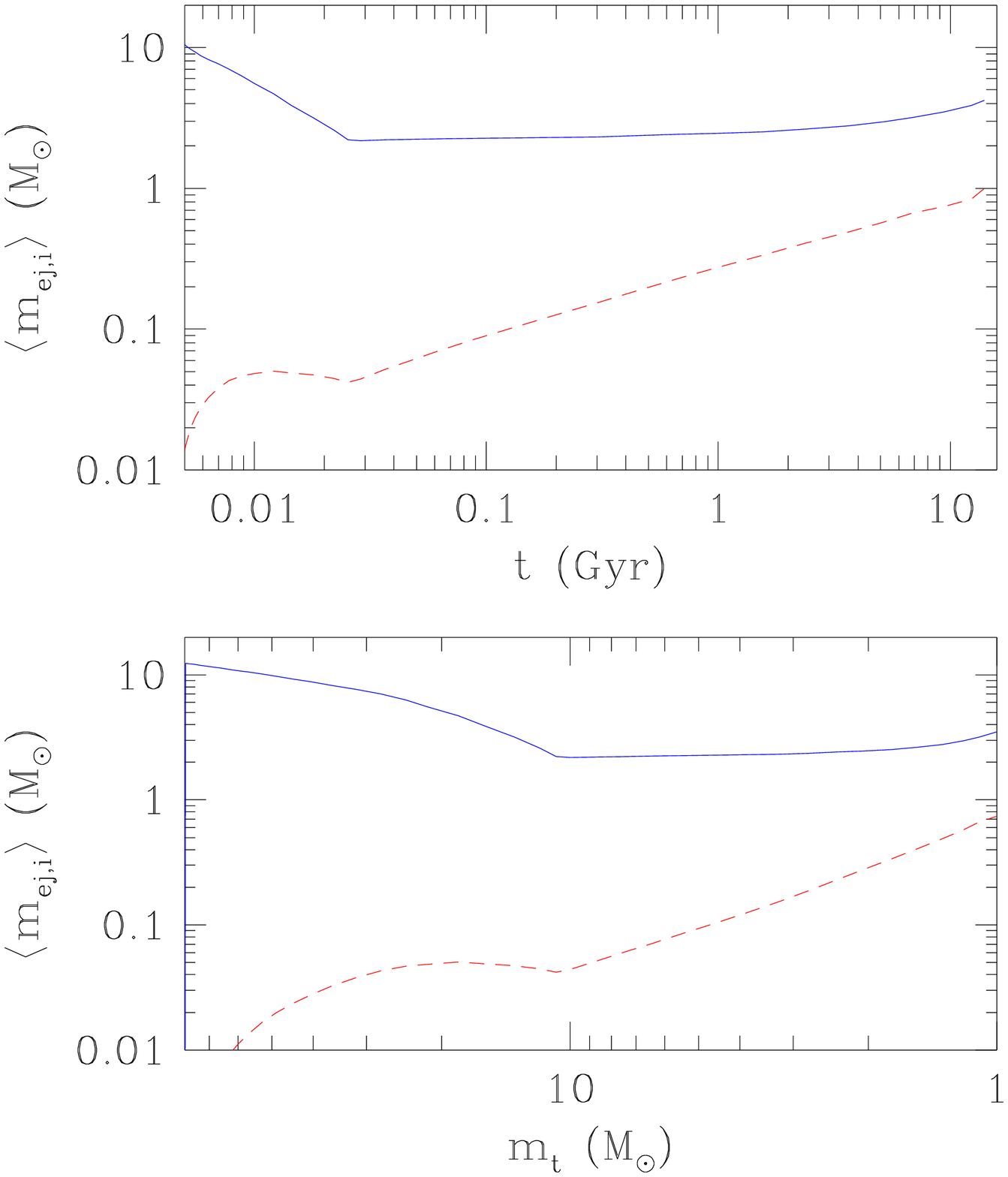}
\end{center}
\caption{}
\end{figure}

\newpage

\begin{figure}[!htb]
\begin{center}
\leavevmode
\epsfxsize=160mm
\epsfbox{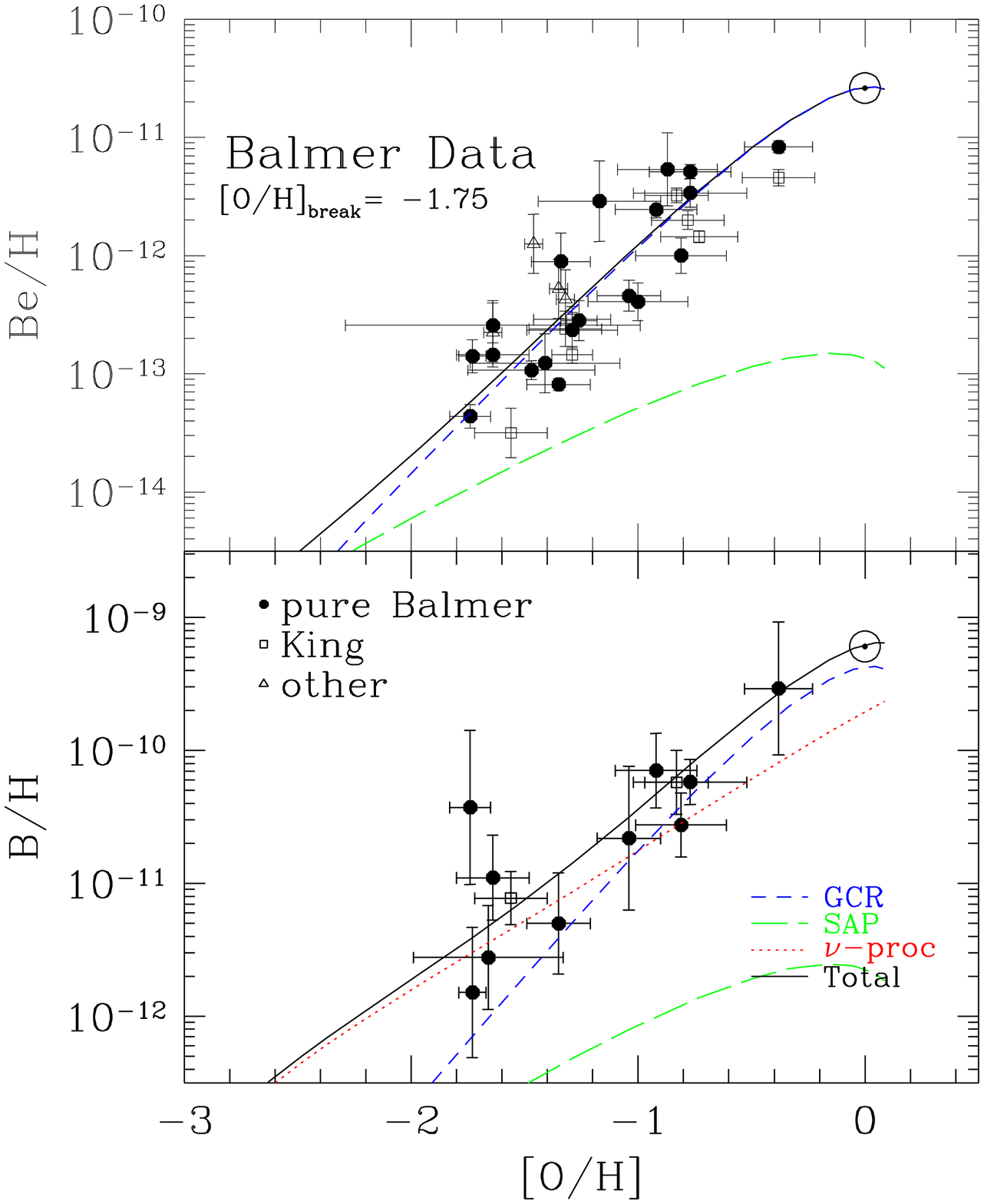}
\end{center}
\caption{}
\end{figure}

\newpage

\begin{figure}[!htb]
\begin{center}
\leavevmode
\epsfxsize=160mm
\epsfbox{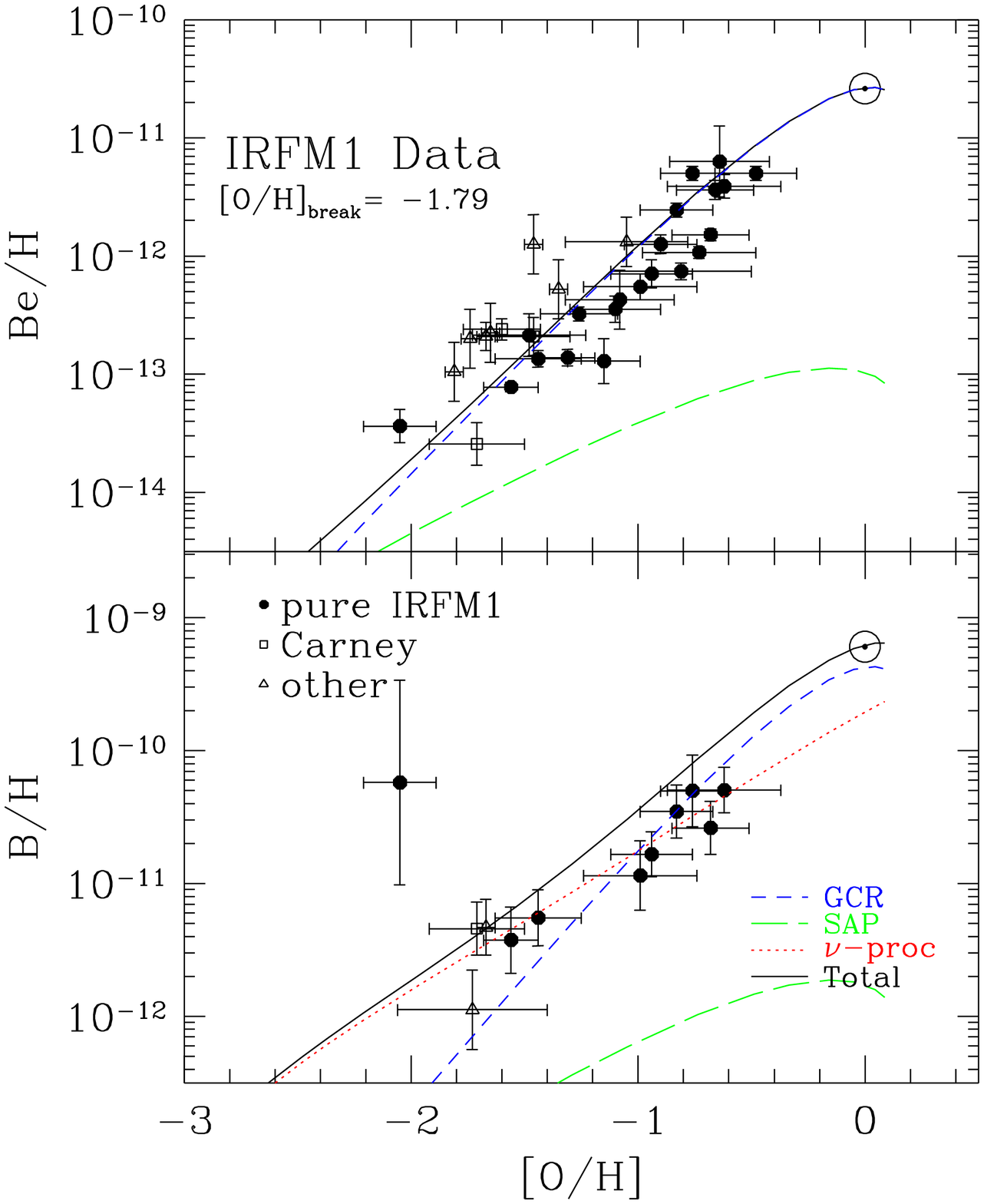}
\end{center}
\caption{}
\end{figure}

\newpage

\begin{figure}[!htb]
\begin{center}
\leavevmode
\epsfxsize=160mm
\epsfbox{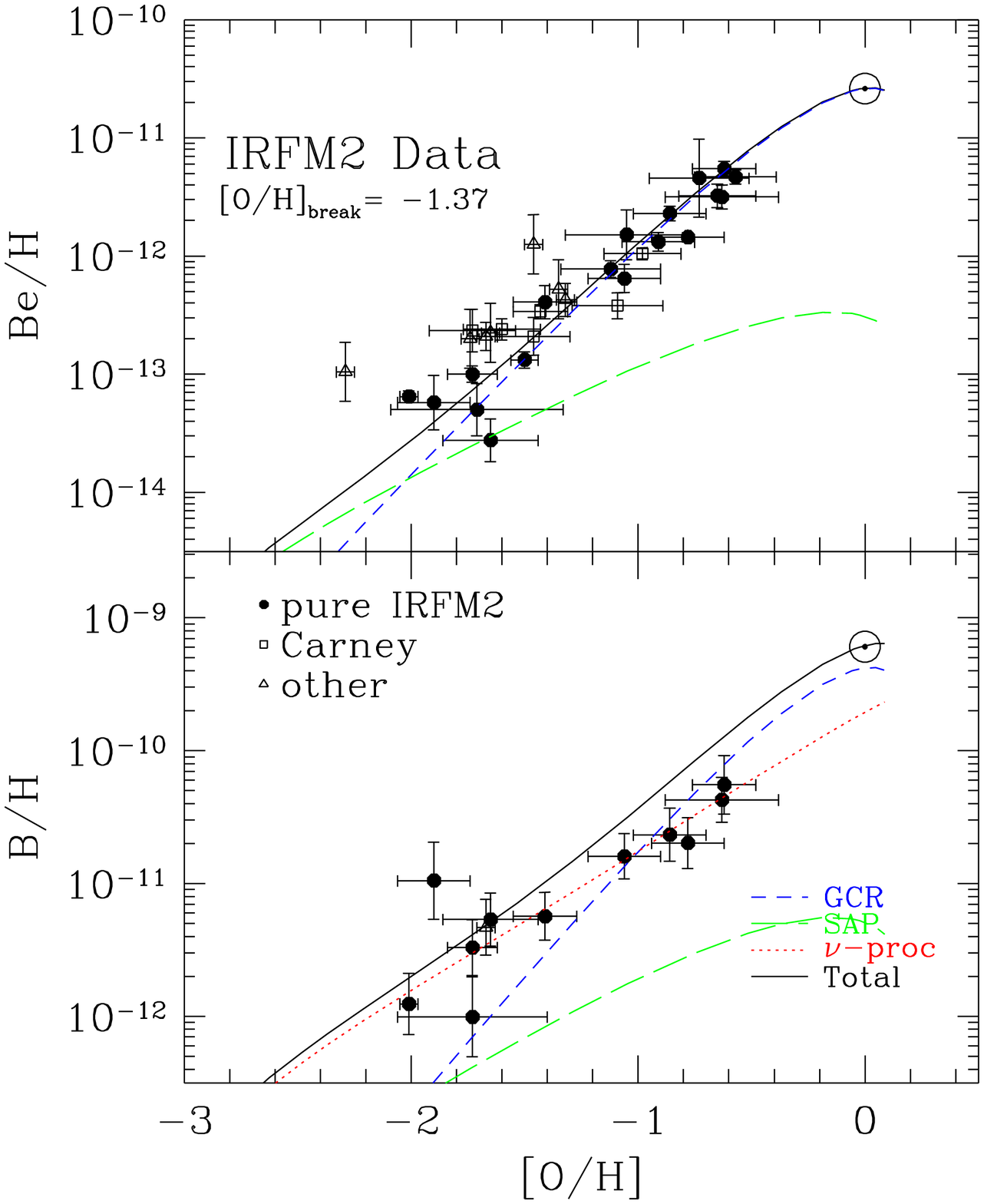}
\end{center}
\caption{}
\end{figure}

\newpage

\begin{figure}[!htb]
\begin{center}
\leavevmode
\epsfxsize=160mm
\epsfbox{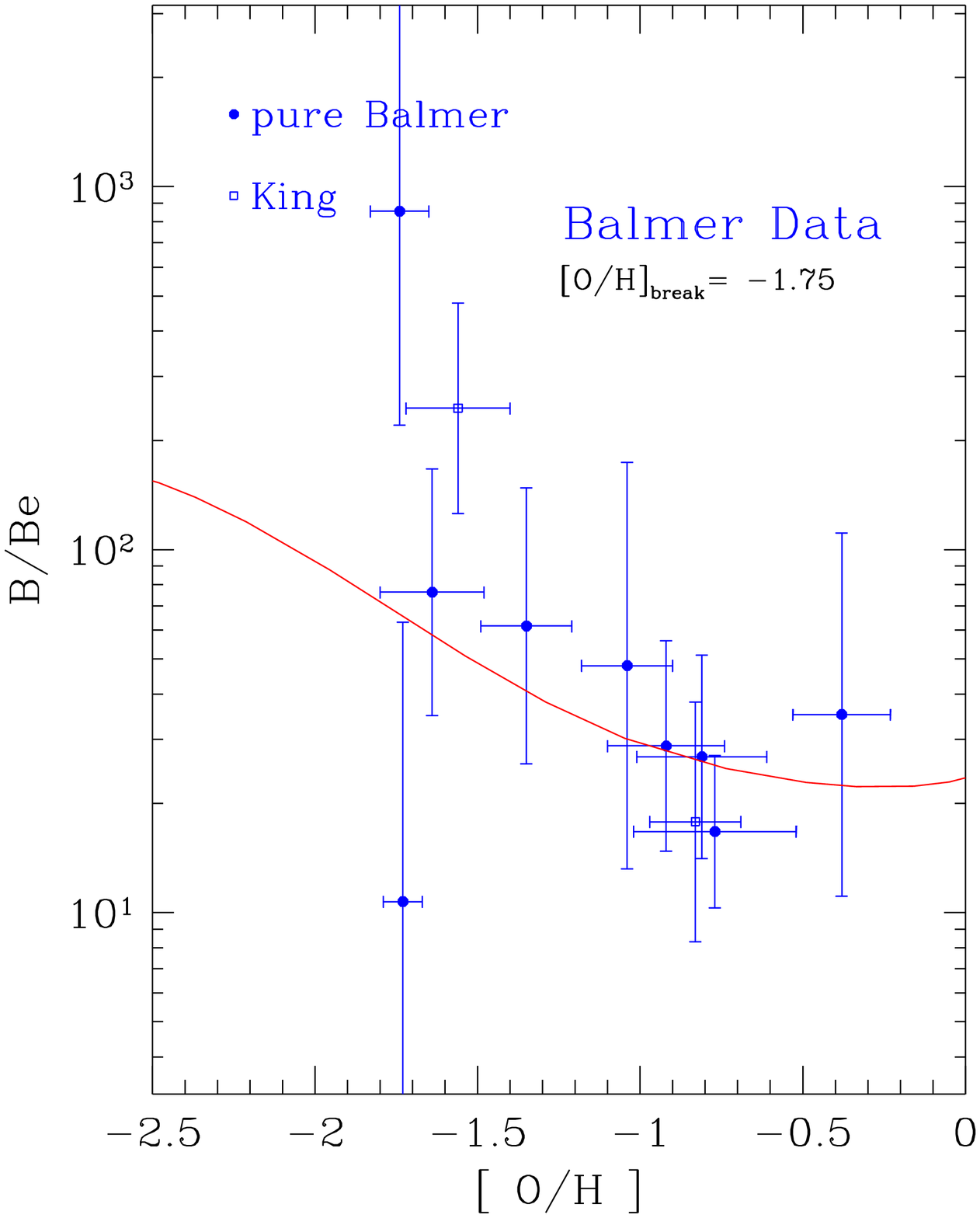}
\end{center}
\caption{}
\end{figure}

\newpage

\begin{figure}[!htb]
\begin{center}
\leavevmode
\epsfxsize=160mm
\epsfbox{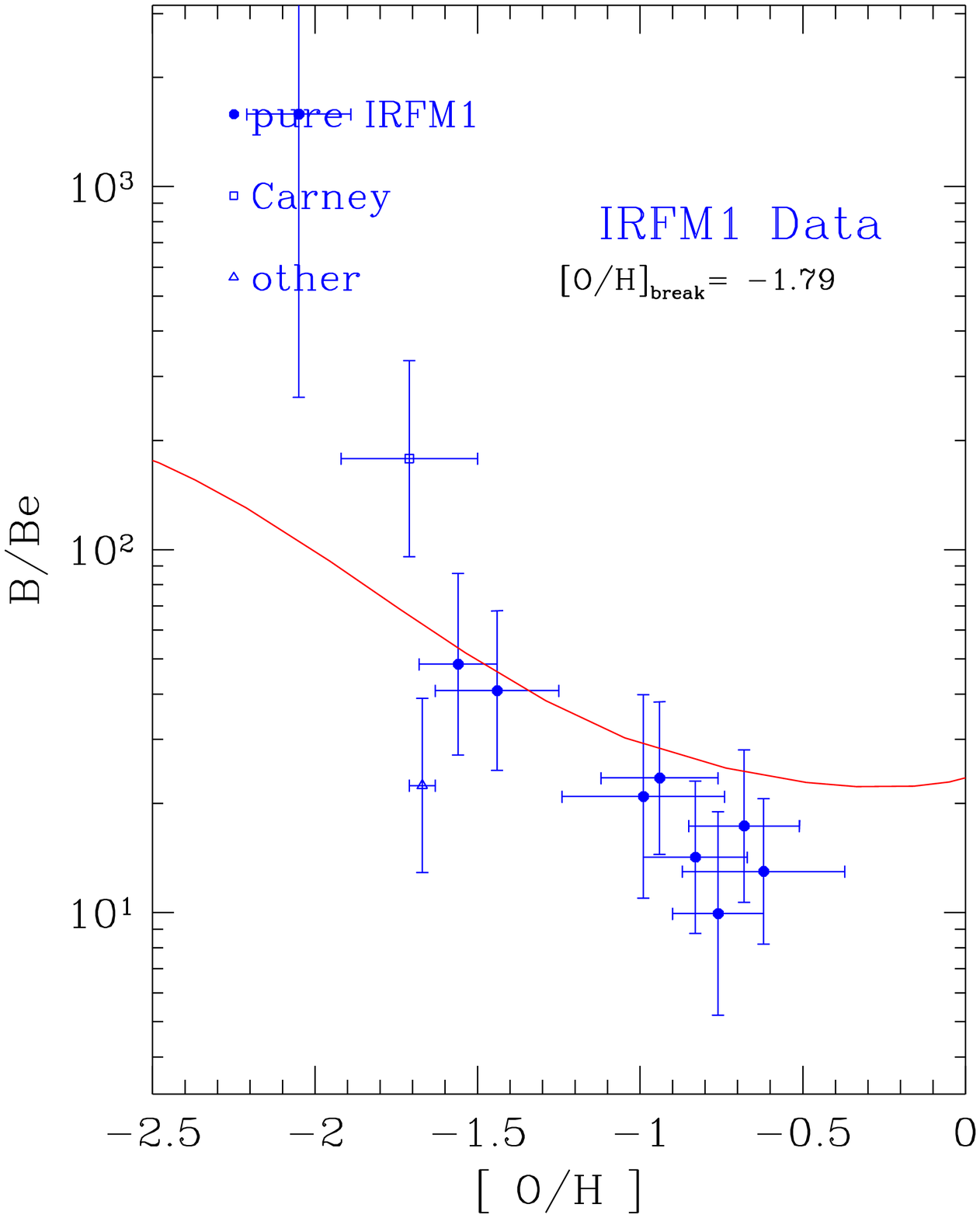}
\end{center}
\caption{}
\end{figure}

\newpage

\begin{figure}[!htb]
\begin{center}
\leavevmode
\epsfxsize=160mm
\epsfbox{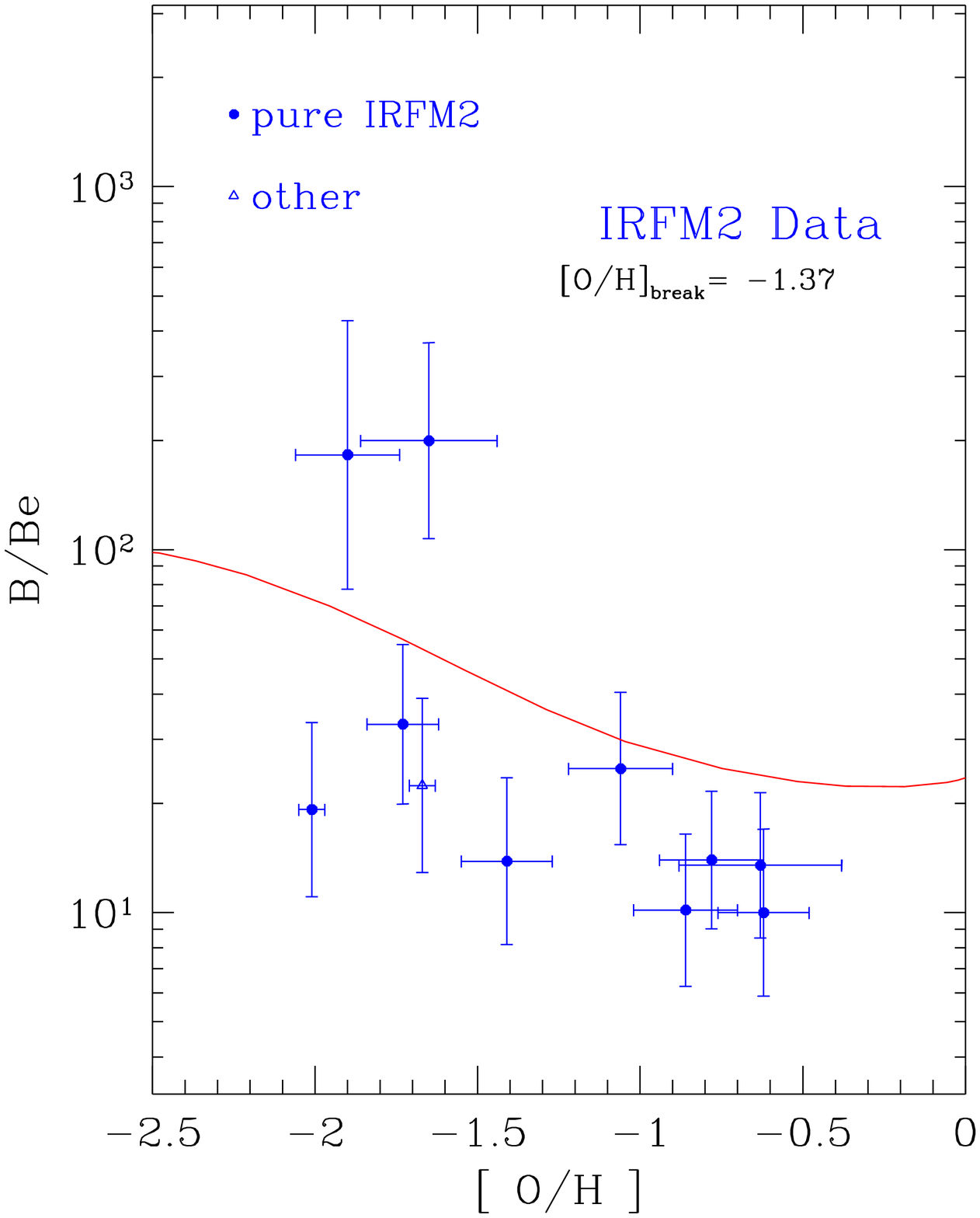}
\end{center}
\caption{}
\end{figure}

\newpage

\begin{figure}[!htb]
\begin{center}
\leavevmode
\epsfxsize=160mm
\epsfbox{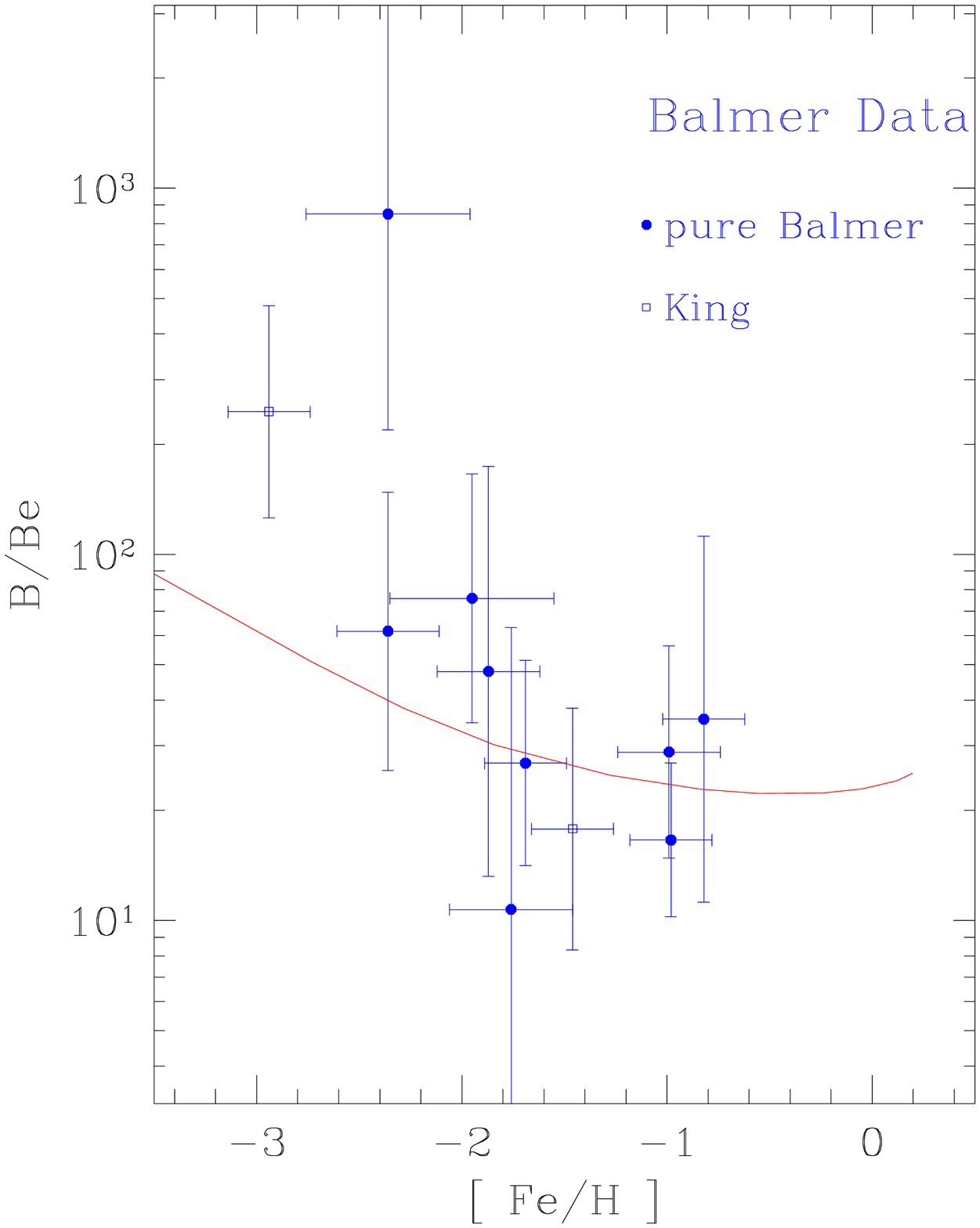}
\end{center}
\caption{}
\end{figure}

\newpage

\begin{figure}[!htb]
\begin{center}
\leavevmode
\epsfxsize=160mm
\epsfbox{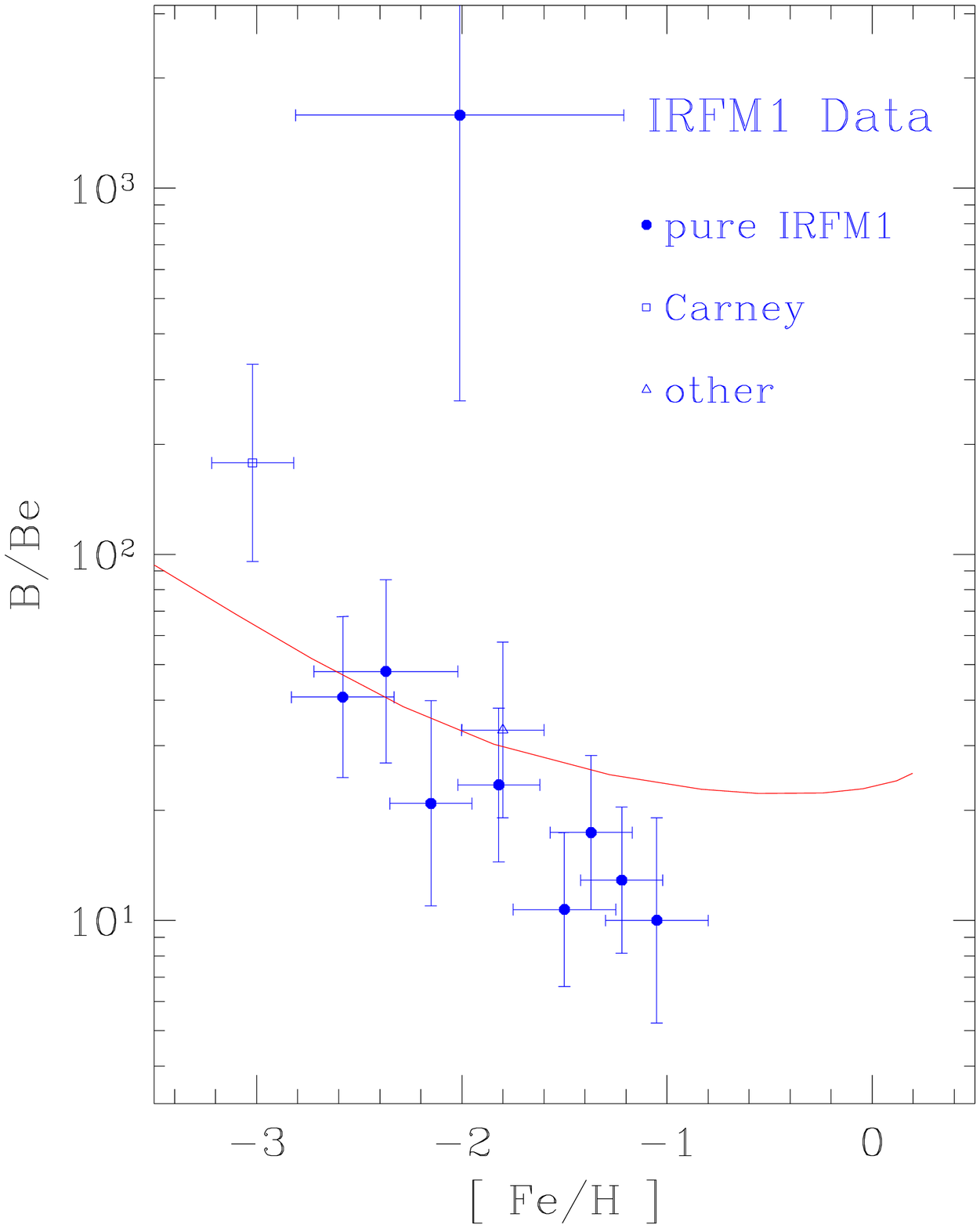}
\end{center}
\caption{}
\end{figure}

\newpage

\begin{figure}[!htb]
\begin{center}
\leavevmode
\epsfxsize=160mm
\epsfbox{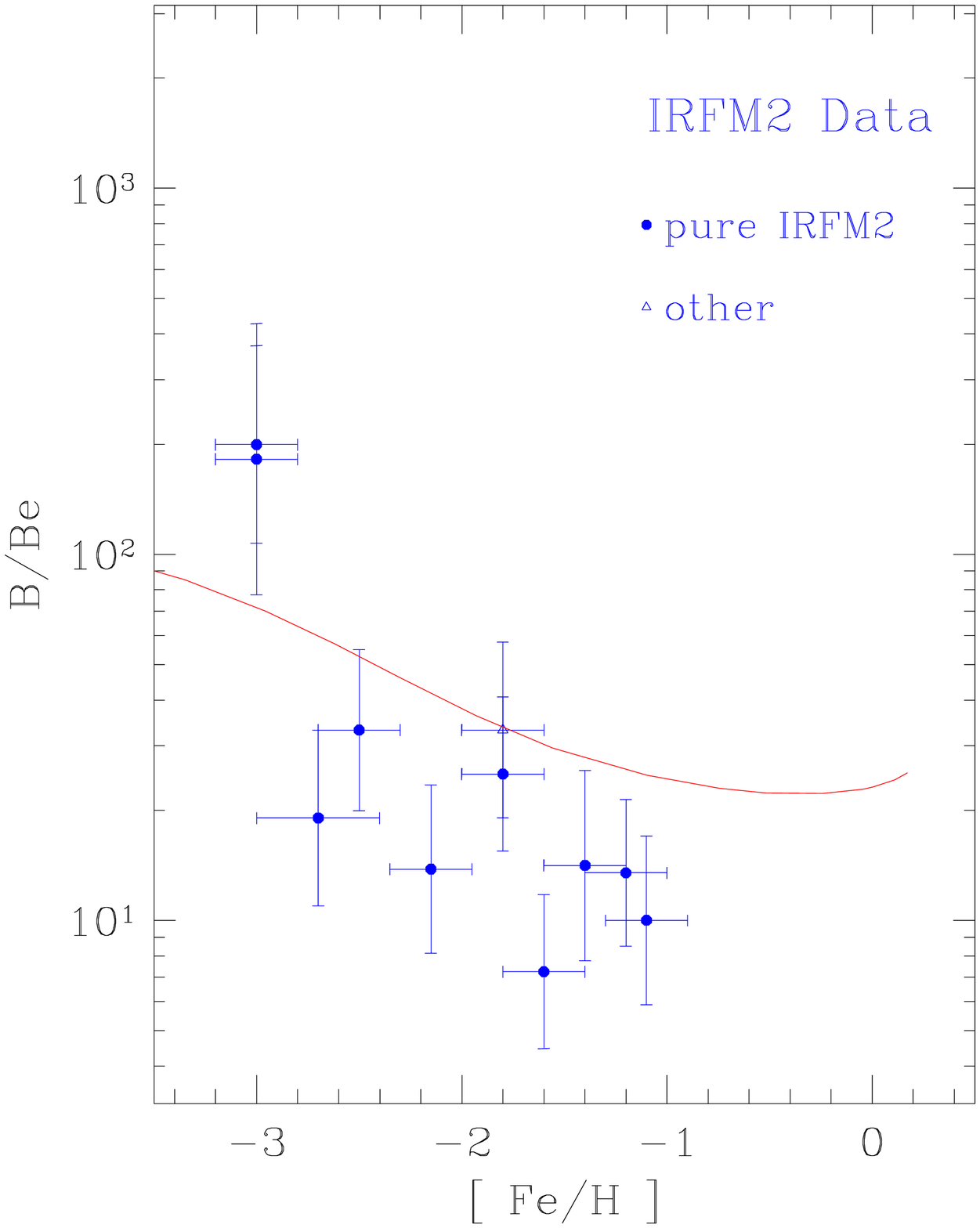}
\end{center}
\caption{}
\end{figure}

\end{document}